\documentclass[aps,pre,citeautoscript,twocolumn,reprint,longbibliography,floatfix,superscriptaddress]{revtex4-2}

\usepackage{hyperref}
\usepackage{xcolor}
\usepackage{graphicx}
\usepackage{amssymb}
\usepackage{amsmath}
\usepackage{mathtools}
\usepackage{epsf}
\usepackage{bm}
\usepackage{cprotect}
\usepackage{comment}
\usepackage{physics}
\usepackage{algorithm}
\usepackage{algorithmic}
\usepackage{caption}
\usepackage{subcaption}
\usepackage{siunitx}

\epsfclipon

\renewcommand{\eqref}[1]{Eq.~(\ref{#1})}

\newcommand{\figref}[1]{Fig.~\ref{#1}}
\newcommand{\figrefs}[1]{Figs.~\ref{#1}}
\newcommand{\tblref}[1]{Table~\ref{#1}}
\newcommand{\secref}[1]{Sec.~\ref{#1}}
\newcommand{\appref}[1]{Appendix~\ref{#1}}
\newcommand{\algref}[1]{Algorithm~\ref{#1}}

\newcommand{\citref}[1]{Ref.~\cite{#1}}
\newcommand{\citrefs}[1]{Refs.~\cite{#1}}
\newcommand{\movref}[1]{Movie~#1}
\newcommand{\movrefs}[1]{Movies~#1}
\newcommand{\etal}{\textit{et al}.\ }
\newcommand{\etals}{\textit{et al}.'s\ }

\newcommand{\dft}[1]{\mathrm{DFT}[#1]}
\newcommand{\idft}[1]{\mathrm{iDFT}[#1]}
\newcommand{\avgx}[1]{\langle{#1}\rangle}

\usepackage{xr}
\makeatletter
\newcommand*{\addFileDependency}[1]{
	\typeout{(#1)}
	\@addtofilelist{#1}
	\IfFileExists{#1}{}{\typeout{No file #1.}}
}
\makeatother

\newcommand*{\myexternaldocument}[1]{
	\externaldocument[S-]{#1}
	\addFileDependency{#1.tex}
	\addFileDependency{#1.aux}
}
\myexternaldocument{suppl}

\begin{document}

\title{Route to turbulence via oscillatory states in polar active fluid under confinement}

\author{Sora Shiratani}
\email{sora.shiratani@phys.s.u-tokyo.ac.jp}
\affiliation{Department of Physics,\! The University of Tokyo,\! 7-3-1 Hongo,\! Bunkyo-ku,\! Tokyo 113-0033,\! Japan}

\author{Kazumasa A. Takeuchi}
\email{kat@kaztake.org}
\affiliation{Department of Physics,\! The University of Tokyo,\! 7-3-1 Hongo,\! Bunkyo-ku,\! Tokyo 113-0033,\! Japan}
\affiliation{Institute for Physics of Intelligence (ipi), The University of Tokyo,
    7-3-1 Hongo, Bunkyo-ku, Tokyo 113-0033, Japan}

\author{Daiki Nishiguchi}
\email{nishiguchi@noneq.phys.s.u-tokyo.ac.jp}
\affiliation{Department of Physics,\! The University of Tokyo,\! 7-3-1 Hongo,\! Bunkyo-ku,\! Tokyo 113-0033,\! Japan}
\affiliation{PRESTO, Japan Science and Technology Agency, 4-1-8 Honcho, Saitama 332-0012, Japan}

\date{\today}

\begin{abstract}
    We report a novel route to active turbulence, observed in numerical simulations of a polar active fluid model under confinement. To deal with large-scale computations with arbitrary geometries, we developed a GPU-based scheme that can be used for any boundary shape in a unified manner. For the circular confinement, as the radius was increased, we found a series of transitions first from a single stationary vortex to an oscillating pair of vortices, then through reentrant transitions between oscillatory and chaotic dynamics before finally reaching the active turbulence. The first transition turned out to be hysteretic, with the emergence of the oscillatory state consistent with the subcritical Hopf bifurcation.
    In dumbbell-shaped boundaries composed of two overlapping circles, we observed a transition comparable to the ferromagnetic-antiferromagnetic vortex-order transition reported in previous experiments, but the transition point turned out to show a qualitatively different geometry dependence.
\end{abstract}

\maketitle

\section{introduction}
Routes to chaos and turbulence have been one of the central topics in fluid mechanics at high Reynolds numbers \cite{landau1987fluid,Eckmann-RMP1981} and statistical physics.
However, turbulent phenomena are widely observed beyond the high-Reynolds-number realm, with a notable example of active turbulence \cite{alert2022active} in low-Reynolds-number active matter systems, for which the route to turbulence remains largely unexplored.
Active matter refers to a collection of self-propelled particles and it often exhibits collective motion due to alignment interaction.
While ordered collective motion often arises in theoretical models \cite{Chate-ARCMP2019} and also occasionally in experiments \cite{Bricard.etal-N2013,Nishiguchi.etal-PRE2017,Iwasawa.etal-PRR2021}, it is not rare that destabilizing interactions also act and render the collective motion turbulent.
Such active turbulence has indeed been observed in various experimental systems, such as reconstituted cytoskeletal systems \cite{sanchez2012spontaneous}, electrokinetic Janus particles \cite{nishiguchi2015mesoscopic}, sperms \cite{creppy2015turbulence} and bronchial epithelial cell cultures \cite{blanch2018turbulent}.
As demonstrated thereby, active turbulence is characterized by collective motion with many swirls and vortices despite the low Reynolds numbers, which has been diagnosed, among other approaches, through scaling behavior of the power spectrum \cite{alert2022active}.
Hydrodynamic descriptions were also proposed, which successfully reproduced dynamics and statistical properties of bulk active turbulence \cite{wensink2012mesoscale, dunkel2013fluid, dunkel2013minimal, alert2022active}.

Besides these developments on bulk systems, the presence of boundaries and confinement have provided interesting new perspectives.
While theoretical approaches are often difficult because of the \textit{a priori} unknown boundary condition for hydrodynamic descriptions and heavy computational costs of agent-based simulations including hydrodynamic effects, experiments have shown that active turbulence often self-organizes into ordered states under confinements.
For example, an ordered vortex has been observed in circular geometries and a directed flow in channels, in diverse systems such as bacterial suspensions \cite{wioland2013confinement,wioland2016directed}, epithelial cells \cite{doxzen2013guidance}, and reconstituted cytoskeletons \cite{wu2017transition,opathalage2019selforganized}.
Pillars were reported to rectify bacterial active turbulence by pinning topological defects in the flow field \cite{nishiguchi2018engineering,reinken2020organizing,figueroa-morales2022}. More elaborated geometries, such as connected circular chambers, were also studied and reported to show transitions between antiferromagnetic and ferromagnetic vortex order \cite{wioland2016ferromagnetic,beppu2017geometry,beppu2021edge}.

It is then natural to ask how ordered active flows under confinements are destabilized as the confinement is weakened and end up in the bulk active turbulence, i.e., the route to active turbulence.
In the literature, routes to active turbulence have been studied more often without boundaries, typically by changing the activity, and different pathways were proposed for different symmetries \cite{alert2022active}:
While active nematic fluids tend to undergo transitions akin to excitable systems \cite{giomi2011excitable}, some compressible polar active fluids show oscillatory phenomena due to self-advection of the polar order \cite{giomi2008complex,giomi2012polar}.
In contrast, studies of confined systems were limited so far to a few cases.
For channels, the transition to active nematic turbulence was numerically studied and reported to be in the directed percolation universality class \cite{doostmohammadi2017onset}, somewhat similarly to shear-driven transitions to turbulence in Navier-Stokes fluids \cite{sano2016universal,Lemoult.etal-NP2016}.
For circular confinements, experiments of reconstituted nematic cytoskeletons showed that topological defects play crucial roles in dynamics and transitions of vortices under strong confinements \cite{opathalage2019selforganized}.
To our knowledge, such routes to turbulence have not been studied so far for polar active systems.

In this paper, we determine the route to turbulence in polar active fluid under varying confinements, using a prototypical model of polar active turbulence known as the Toner-Tu-Swift-Hohenberg (TTSH) equation \cite{wensink2012mesoscale, dunkel2013fluid, dunkel2013minimal, alert2022active} and the boundary conditions identified by previous experiments on bacterial turbulence \cite{reinken2020organizing}.
By increasing the radius of the circular confinement, we found a series of transitions starting from a single stationary vortex, passing through intermediate oscillatory states and finally reaching the active turbulence.
The first transition is from a single stationary vortex to a periodically oscillating pair of vortices and turned out to be hysteretic.
This is followed by a sequence of transitions across periodic and chaotic oscillations, as well as a quasiperiodic one, which consists of two fundamental frequencies with an irrational ratio, before reaching the final turbulent state.
This anomalous route to turbulence is a novel scenario different from those known for the conventional Navier-Stokes turbulence \cite{landau1987fluid,Eckmann-RMP1981} and for active nematics in a channel \cite{doostmohammadi2017onset}.
This finding was made possible by a GPU-based solver that we developed here, which efficiently integrates the TTSH equation for arbitrary boundary shapes.
GPU implementation allowed us to achieve high-resolution computation within reasonable time, which was essential to correctly identify intermediate oscillatory states and transitions.
It also helped us to characterize detailed properties of the transitions, in particular the Lyapunov exponents, which require massive extra computations.
As our scheme works with arbitrary boundary configurations in a unified manner, we expect it to be a useful platform to predict behavior of active turbulence under confinement, which can also be used for designing experimental setups for bacterial turbulence.

This paper is organized as follows.
In \secref{sec:method}, we describe our calculation scheme, regarding how to fully automate calculations under arbitrary-shaped geometries, possible artifacts caused by ordinary schemes, and advantages of GPU implementation.
Then, the results are presented in two sections.
Section~\ref{sec:circle} is for circular confinements, where we unveil the hysteretic stationary-oscillatory transition as well as subsequent transitions between oscillatory and chaotic states.
Section~\ref{sec:dumbbell} reports simulation results for dumbbell-shaped boundaries, which consist of two overlapping circles, chosen here as a test case of the TTSH equation with complex boundary shapes.
In this geometry, we found a transition comparable to the ferromagnetic-antiferromagnetic vortex-order transition reported by previous experiments \cite{beppu2017geometry,beppu2021edge}, but the transition point turned out to show a qualitatively different geometry dependence.
Section~\ref{sec:discussion} is devoted to discussions.
Finally, we summarize the results and give concluding remarks in \secref{sec:conclusion}.

\section{method}  \label{sec:method}
\subsection{model}
We use the TTSH equation, which describes bulk behavior of turbulent collective motion spontaneously formed in dense bacterial suspensions \cite{wensink2012mesoscale, dunkel2013fluid, dunkel2013minimal, alert2022active, reinken2020organizing}.
In its dimensionless form \cite{reinken2020organizing}, it reads:
\begin{gather}
    \div{\vb{v}} = 0, \label{eq:orig1} \\
    \pdv{\vb{v}}{t} + \lambda \vb{v} \vdot \grad{\vb{v}} = a \vb{v} - b \vb{v}^2 \vb{v} - \qty(1 + \laplacian)^2 \vb{v} - \grad{p}, \label{eq:orig2}
\end{gather}
where $\vb{v}(\vb{x},\ t)$ is a coarse-grained velocity field of bacteria, $p$ is an effective pressure that ensures the incompressibility, and $\lambda,\ a,\ b$ are model parameters that do not depend on $\vb{x}$, $t$, $\vb{v}$ or $p$.
Here, following \citref{reinken2020organizing}, we express the dimensionless coordinates $\vb{x}$ and time $t$ in the unit of the characteristic length and time scales, respectively, determined by the Swift-Hohenberg-like term $\qty(1 + \laplacian)^2 \vb{v}$, which has coefficients otherwise.
As a result, the characteristic wavenumber $|\vb{k}_{*}|$ excited by the Swift-Hohenberg-like term is equal to unity, so that the typical size of vortices is $2\pi$ in real space. This corresponds to $\approx 105$~\si{\micro\metre} for bacterial turbulence of {\it Bacillus subtilis} \cite{reinken2020organizing}.
In the present work, we focus on two-dimensional systems, to make our results comparable with experimental observations.

The boundary conditions for the TTSH equation were determined experimentally in \citref{reinken2020organizing}, to be $\vb{v} \equiv (v_x,\ v_y)=0$ and the vorticity $\omega \equiv \partial_x v_y - \partial_y v_x =0$, at least for the experimental condition employed therein.
Then, the existence of boundaries (as well as obstacles) can be incorporated by adding damping terms into the TTSH equation in the form of the vorticity equation \cite{reinken2020organizing}:
\begin{gather}
    \div{\vb{v}} = 0, \label{eq:mod1} \\
    \begin{multlined}
        \pdv{\omega}{t} + \lambda \vb{v} \vdot \grad{\omega} = a \omega - b \curl{\qty[\vb{v}^2 \vb{v}]} \\
        - \qty(1 + \laplacian)^2 \omega - \gamma_{\vb{v}} \curl{\qty[K\qty(\vb{x})\vb{v}]} - \gamma_{\omega} K \qty(\vb{x}) \omega, \label{eq:mod2}
    \end{multlined}
\end{gather}
where $K \qty(\vb{x})$ is a nonnegative scalar field such that $K \qty(\vb{x})\approx 0$ inside the system and $K \qty(\vb{x})\approx 1$ outside, and $\gamma_{\vb{v}},\ \gamma_\omega$ are positive parameters representing the damping strengths.
The first damping term in \eqref{eq:mod2} amounts to adding the damping $-\gamma_{\vb{v}} K \qty(\vb{x}) \vb{v}$ to \eqref{eq:orig2} to prevent $\vb{v}$ from growing in the area outside the system ($K \approx 1$, referred to as masked area) without affecting the area inside the system ($K \approx 0$, unmasked area).
The second damping term $- \gamma_{\omega} K \qty(\vb{x}) \omega$ does the same for the vorticity.
In the following, we set $\qty(a,\ b,\ \lambda,\ \gamma_{\vb{v}},\ \gamma_{\omega}) = \qty(0.5,\ 1.6,\ 9,\ 40,\ 4)$, which had been reported to quantitatively reproduce the experimental results in \citref{reinken2020organizing} (see \tblref{tab:param} in \appref{sec:notation} for the list of the parameter values used in this paper).

\begin{figure*}[tb]
    \begin{minipage}[b]{0.31\linewidth}
        \centering
        \subcaption{} \label{fig:emoji_mask}
        \includegraphics[width = \hsize]{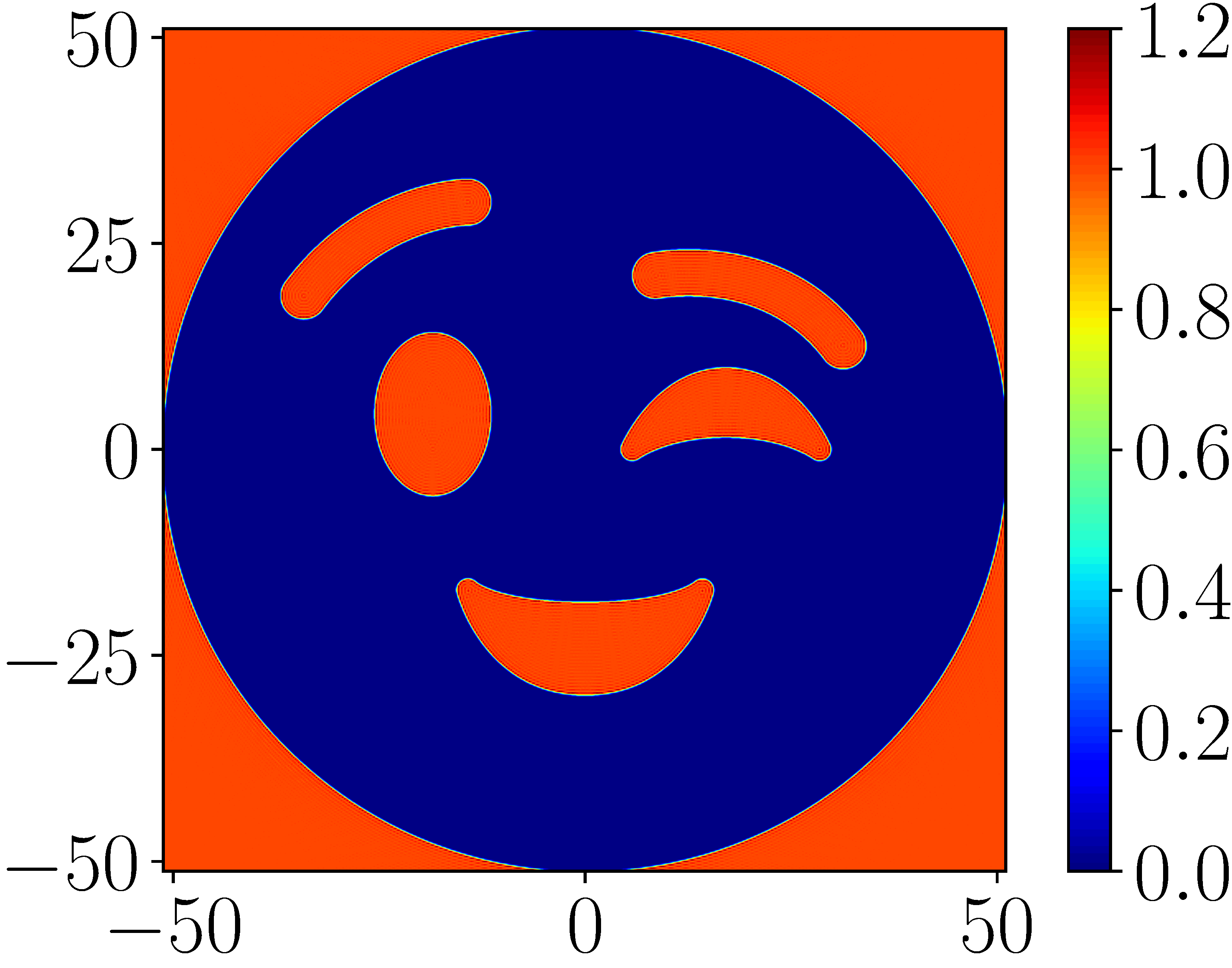}
    \end{minipage}
    \begin{minipage}[b]{0.33\linewidth}
        \centering
        \subcaption{} \label{fig:emoji_snap}
        \includegraphics[width = \hsize]{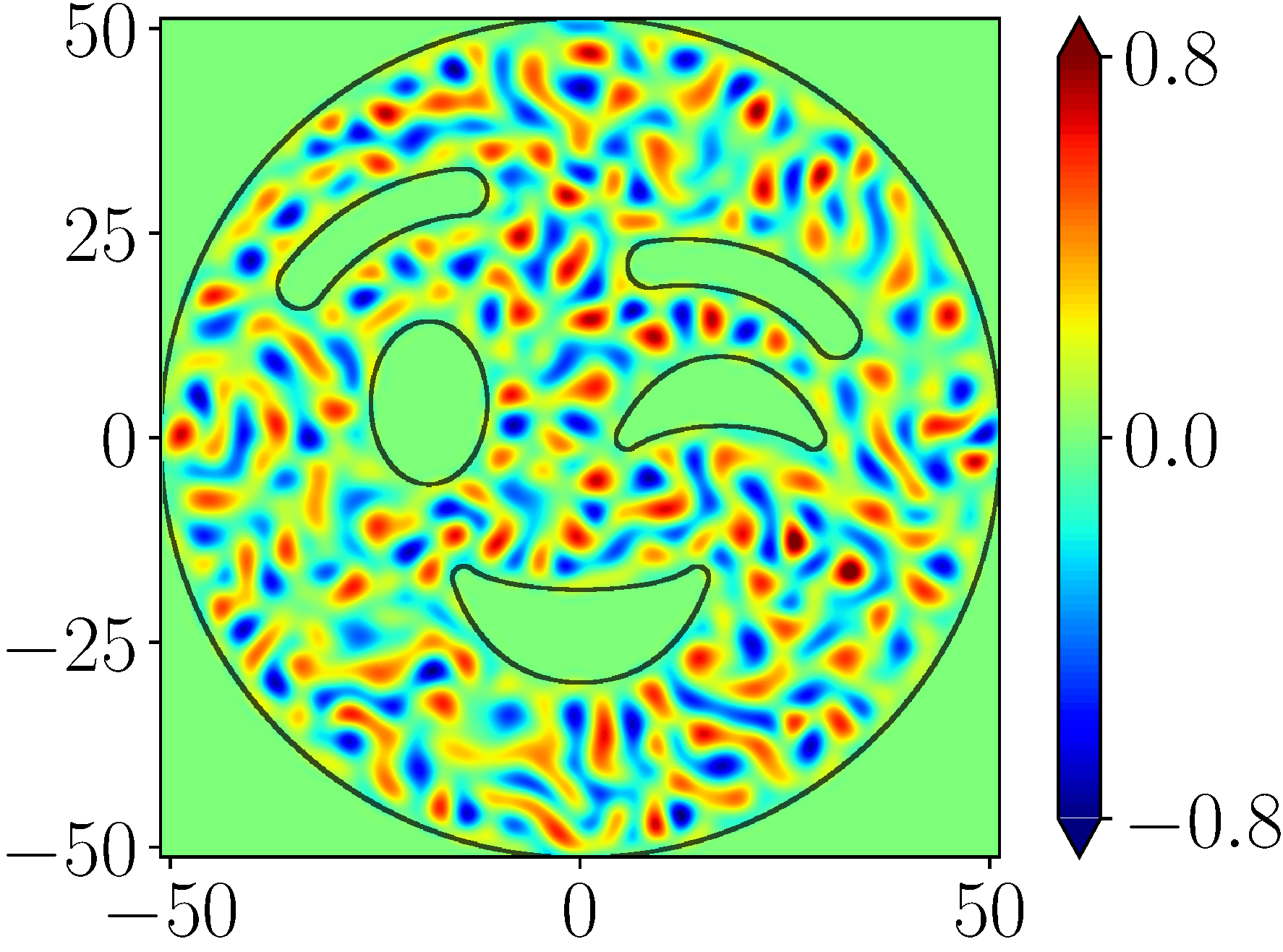}
    \end{minipage}
    \begin{minipage}[b]{0.32\linewidth}
        \centering
        \subcaption{} \label{fig:emoji_mask_slice}
        \includegraphics[width = \hsize]{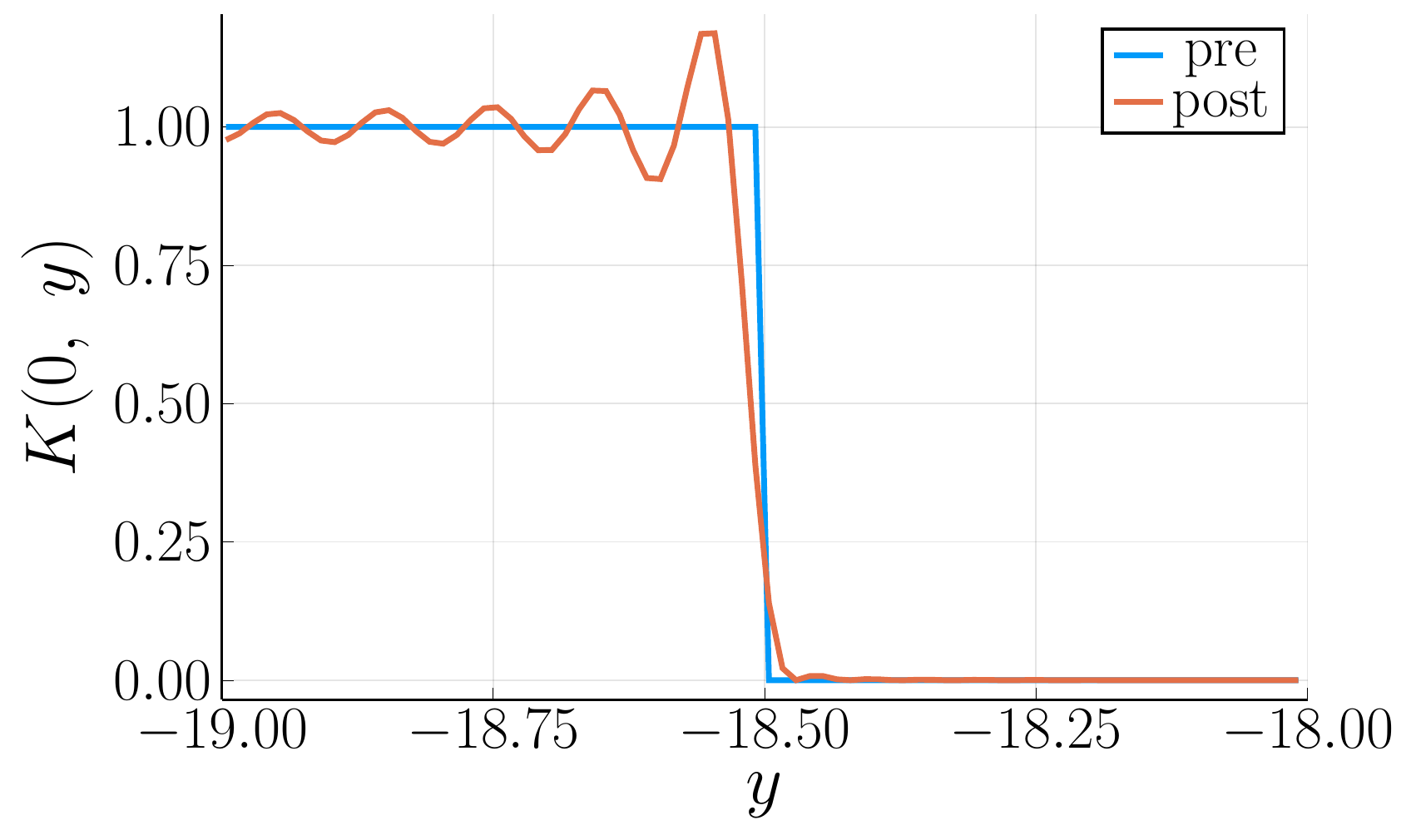}
    \end{minipage}
    \caption{Emoji boundary.
        \subref{fig:emoji_mask} The mask $K(\vb{x})$ generated by our automatic mask generator.
        \subref{fig:emoji_snap} A snapshot of the vorticity field from a simulation using the generated mask. Note that the color scale displayed here is used in all vorticity snapshots in the present paper. See also \movref{S1} \cite{SM}.
        \subref{fig:emoji_mask_slice} Emoji mask before and after the low-pass filtering in the automatic mask generator, sliced at $x = 0$.}    \label{fig:emoji}
\end{figure*}

\subsection{computation}
Although we use the standard pseudospectral method to integrate the TTSH equation, the calculation is not an easy task because large-scale computations are required by nature. There are two specific reasons for the difficulty. First, as previously stated, we use the virtual boundaries represented by the damping terms. This method is sometimes problematic because $\vb{v}$ and $\omega$ penetrate into the damped area outside the system. To avoid artifacts, damping walls need to be sufficiently thick so that the penetrating fields decay to almost zero. Second, the spatial discretization interval $\Delta x$ needs to be small because the results may be influenced by small changes in the mask field $K(\vb{x})$ as addressed in the next section. Fortunately, this model is not so sensitive to the temporal discretization interval $\Delta t$ if implemented properly.
This allowed us to use the Euler method and obtain converging results efficiently.
Detailed descriptions of the algorithm are provided in \appref{sec:scheme}.
Care was taken to make full use of the performance of GPU (see \appref{sec:GPU}).
Taking these properties into account, we discretized the model on an $8192 \times 8192$ square lattice with $\Delta x = 0.005$ and simulated its time evolution with $\Delta t = 0.01$, unless otherwise stated. Note that most of the calculation was done for the masked region (typically $~90 \ \%$ in area), to construct a thick barricade to ensure the decay of the penetrating fields.

\subsection{mask generation}
Although the physical meaning of the mask field $K(\vb{x})$ is straightforward, due care should be taken to design a suitable $K(\vb{x})$ for computation, due to constraints originating from the pseudospectral method (see \appref{sec:maskgen}).
One such constraint is the $\frac{1}{2}$ rule for antialiasing in the case of the cubic nonlinearity (see \appref{sec:scheme}). In short, $K(\vb{x})$ should not contain high-wavenumber modes in order to properly integrate the nonlinear terms of the TTSH equation, and therefore a naive step-function-like binary mask cannot be used. An intuitive treatment is to interpolate the gap between $K = 0$ (unmasked) and $K = 1$ (masked) by a slowly-varying function such as $\tanh$, but this is not an ideal solution because $\tanh$ contains high-wavenumber modes even though their amplitudes are significantly smaller than the original binary mask.
Moreover, this approach requires us to design and encode the mask explicitly as a part of the program before the calculation. This task is not straightforward when the geometry is complex, and there is no general way of predicting whether aliasing noise associated with the mask is actually tolerable or not.
Therefore, even though the TTSH equation has been actually calculated by using such a mask without breaking down, possibly thanks to the Swift-Hohenberg-like term $\qty(1 + \laplacian)^2 \vb{v}$ which damps high-wavenumber modes, it is desirable to design and use a mask that is free of any aliasing noise.

To overcome these difficulties, we adopted a completely different approach, which automatically generates a mask $K(\vb{x})$ that strictly satisfies all the requirements by using a low-pass filter (see \appref{sec:maskgen} for details). As a result, the only required input is a binary scalar field indicating where to mask.
Figure~\ref{fig:emoji} displays a toy example using a binarized unicode emoji \cite{emoji} as the boundary (see also \movref{S1} \cite{SM}).
As a result of the low-pass filtering, the generated mask $K(\vb{x})$ oscillates as displayed in \figref{fig:emoji_mask_slice}.
Although this side effect is more or less inevitable, the oscillation can be suppressed by introducing additional low-pass filtering operations, at the price of sharpness.
We, however, did not take this option intentionally because we prioritized the sharpness and we did not see any discernible artifact.

\section{confinement in circle}  \label{sec:circle}
\label{sec:ConfinementInCircle}
\subsection{motivations and background}
Circle is one of the simplest geometries parametrized only by its radius $R$, and its simplicity has long been arousing experimental interests as surveyed in the introduction section.
In particular, it was reported that bacterial flow can be stabilized into a single vortex if confined in a sufficiently small circular geometry\cite{wioland2013confinement,beppu2017geometry,beppu2021edge}, while the opposite, unconfined limit corresponds to turbulence.
This led us to perform numerical simulations in circular areas for various radii $R$ and investigated the route to active turbulence in this case.
Unless otherwise stated, we started from a random initial state and discarded transients to ensure that the system is in a steady state.
Note that high-resolution calculation realized by GPU is crucial here (see \appref{sec:GPU}), because it turned out that small changes in $R$ may affect qualitative features of the flow.

\begin{figure*}[tb]
    \begin{minipage}[b]{\linewidth}
        \centering
        \subcaption{} \label{fig:phase}
        \includegraphics[width = \hsize]{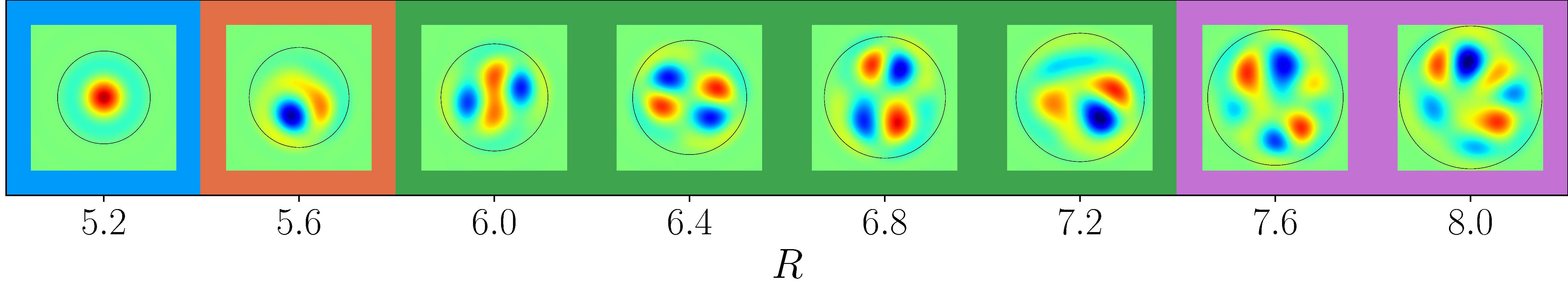}
    \end{minipage}
    \begin{minipage}[b]{\linewidth}
        \centering
        \subcaption{} \label{fig:anime}
        \includegraphics[width = \hsize]{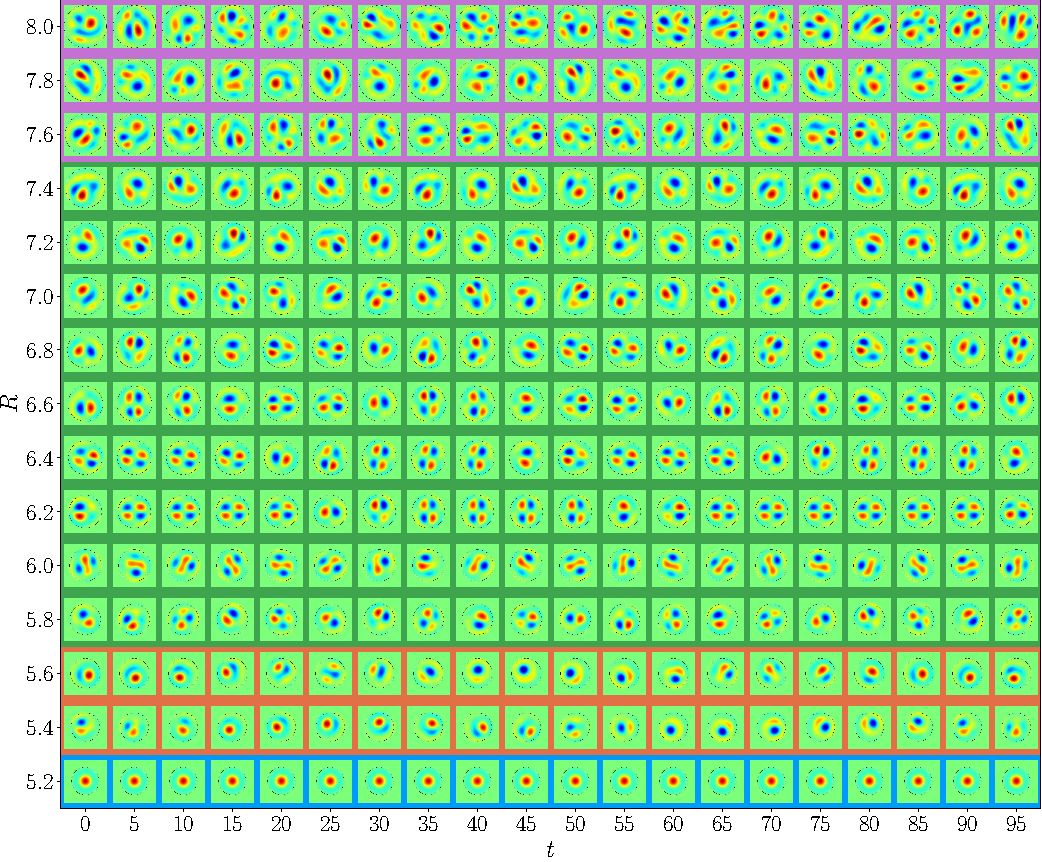}
    \end{minipage}
    \centering
    \caption{
        Phase diagram for the circular confinement, showing snapshots of the vorticity field for different $R$ (same color scale as \figref{fig:emoji_snap}).
        The frames of the snapshots are given different colors corresponding to the number of vortices (blue: one, orange: two, green: four, purple: more).
        \subref{fig:phase} Typical snapshot chosen for each $R$.
        \subref{fig:anime} Time series of the vorticity field. See also \movref{S2} \cite{SM}.
    } \label{fig:phase_anime}
\end{figure*}

\begin{figure}[tb]
    \centering
    \includegraphics[width = \hsize]{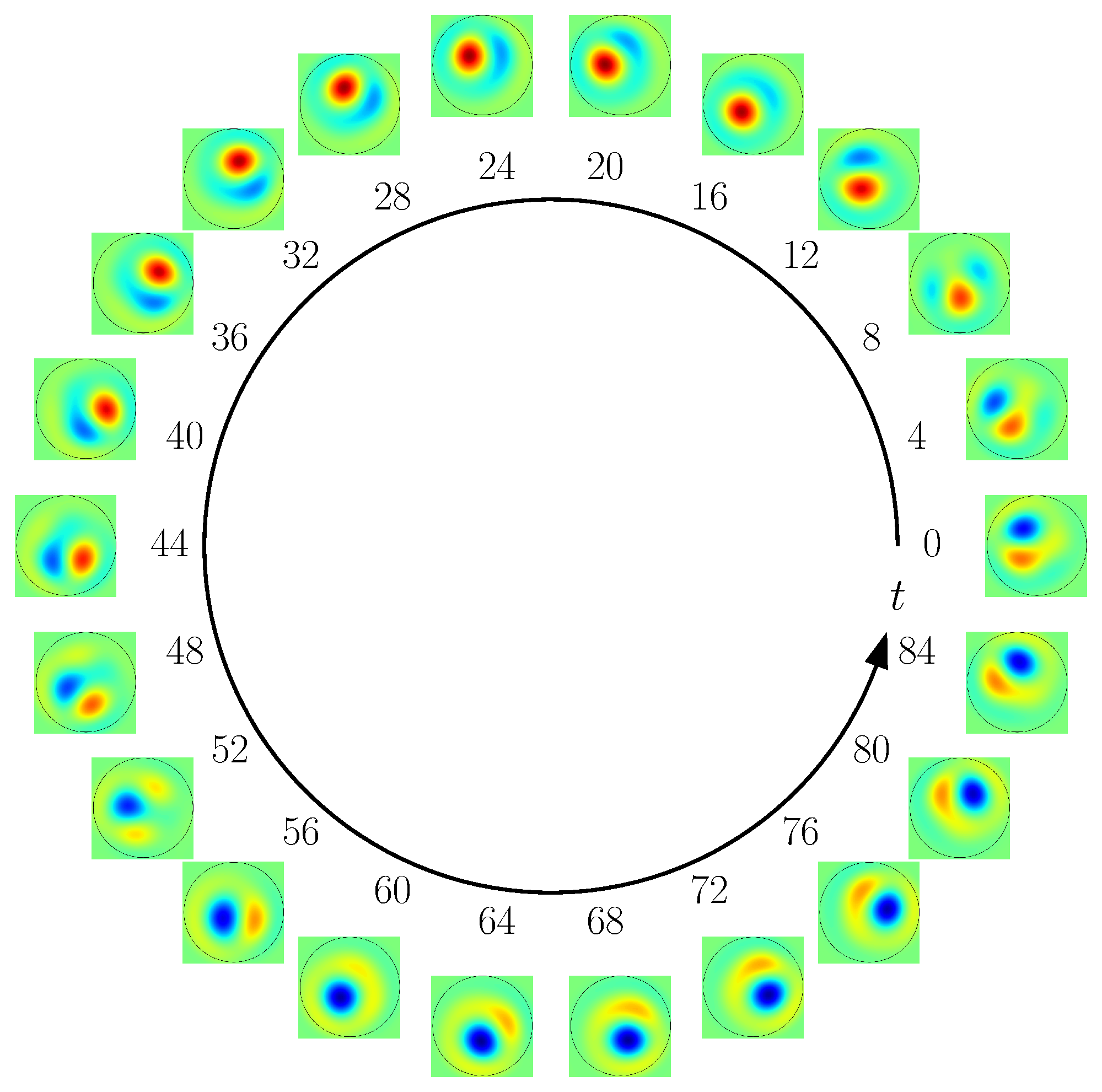}
    \caption{
        Time series of the vorticity field in the oscillating vortex pair state at $R = 5.4$. Same color scale as \figref{fig:emoji_snap}.
        See also \movref{S4} \cite{SM}.
    } \label{fig:ring}
\end{figure}

\subsection{overview of results} \label{sec:overview}
First we outline the results obtained in this geometry.
Our calculations revealed a rich phase diagram with different numbers of vortices [\figref{fig:phase}] and different dynamical states [\figref{fig:anime} and \movrefs{S2-S8} \cite{SM}] depending on the radius $R$.
While no vortex is observed for $R \lesssim 5.1$ due to strong damping, around $R = 5.2$ we observe a single stationary vortex generated at the center of the circle (blue box in \figref{fig:phase_anime} and \movref{S3} \cite{SM}).
As $R$ is increased, the number of vortices increases to two (orange box), four (green), and more (purple).
Dynamics becomes non-stationary as soon as multiple vortices are generated.
In particular, the first non-stationary state observed around $R = 5.4$ consists of an oscillating pair of vortices, as shown in \figref{fig:ring} and \movref{S4} \cite{SM}.
The transition between the single-vortex stationary state and the vortex-pair oscillatory state will be characterized in detail, in \secref{sec:hysteresis}.

\begin{figure}[tb]
    \centering
    \includegraphics[width = \hsize]{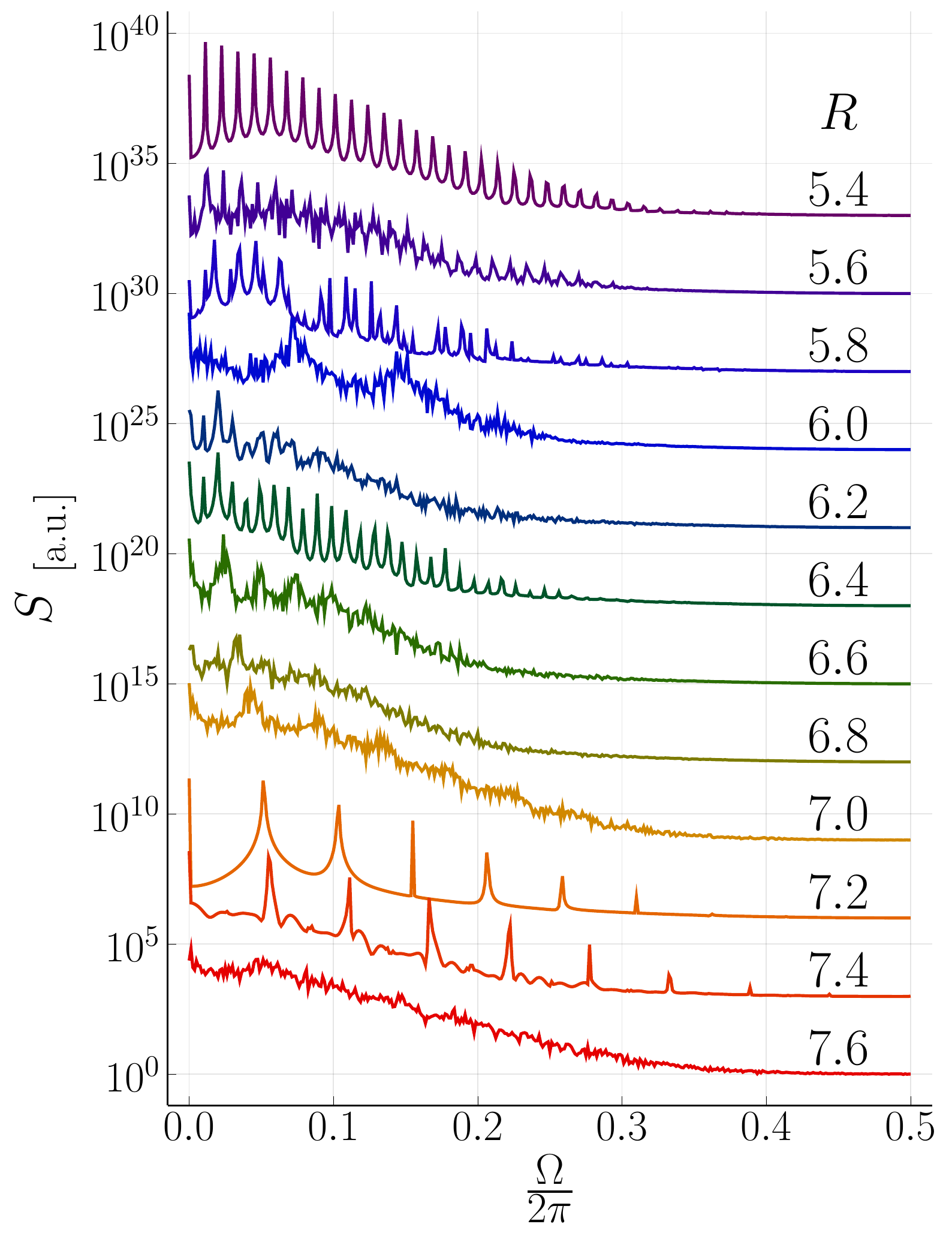}
    \caption{
        Temporal PSD of the vorticity field [\eqref{eq:PSD}] measured for different $R$.
        Each PSD is multiplied by a constant to avoid overlapping.
    } \label{fig:fft}
\end{figure}

To characterize the changes in the dynamics more quantitatively, we measure the temporal power spectral density (PSD) of the vorticity field (\figref{fig:fft}), defined by
\begin{equation}
    S(\Omega) \equiv \frac{1}{N^2} \sum_{\vb{x}} \qty|\sum_{t} \omega\qty(\vb{x},\ t) e^{-i \Omega t}|^2.  \label{eq:PSD}
\end{equation}
At $R = 5.4$ (\movref{S4} \cite{SM}), we confirm the periodic oscillation by the PSD that consists of a single fundamental frequency and its harmonics. At $R = 5.6$ (\movref{S5} \cite{SM}), the first chaotic region suddenly appears while the oscillatory dynamics is still observed. At $R = 5.8$ (\movref{S6} \cite{SM}), the system regains regularity but another fundamental frequency emerges and the oscillation becomes quasiperiodic (with the fundamental frequencies being $0.01125$ and $0.0175$).
Simultaneously, the system now accommodates four vortices (see \figref{fig:phase_anime}, \movref{S2} \cite{SM}).
From $R = 6.0$ to $R = 7.4$, the system goes back and forth between the chaotic and oscillatory states (see \movref{S7} \cite{SM} for $R=7.2$) until it finally falls into the chaotic state at $R = 7.6$ (\movref{S8} \cite{SM}) and never returns.
From $R = 7.6$, the system has more than four vortices and the number increases with $R$.

In the following, we will characterize the single-vortex stationary state (\secref{sec:phaseI}) and the transition to the vortex-pair oscillatory state (\secref{sec:hysteresis}).
Then we will comment on the route to turbulence observed in this geometry (\secref{sec:route}).

\subsection{single-vortex stationary state}\label{sec:phaseI}

In this state observed around $R = 5.2$ (blue box in \figref{fig:phase_anime}), a single vortex is generated and located at the center of the circle. The sign of the vorticity is determined by the initial condition that we generated at random. This is reminiscent of the single-vortex stationary state reported in experiments \cite{wioland2013confinement,enkeleida2014fluid,beppu2017geometry,beppu2021edge}, even though the boundary condition may differ and no edge currents are observed in our simulations.

\begin{figure}[t!]
    \begin{minipage}[b]{0.48\linewidth}
        \centering
        \subcaption{
        } \label{fig:slice}
        \includegraphics[width = \hsize]{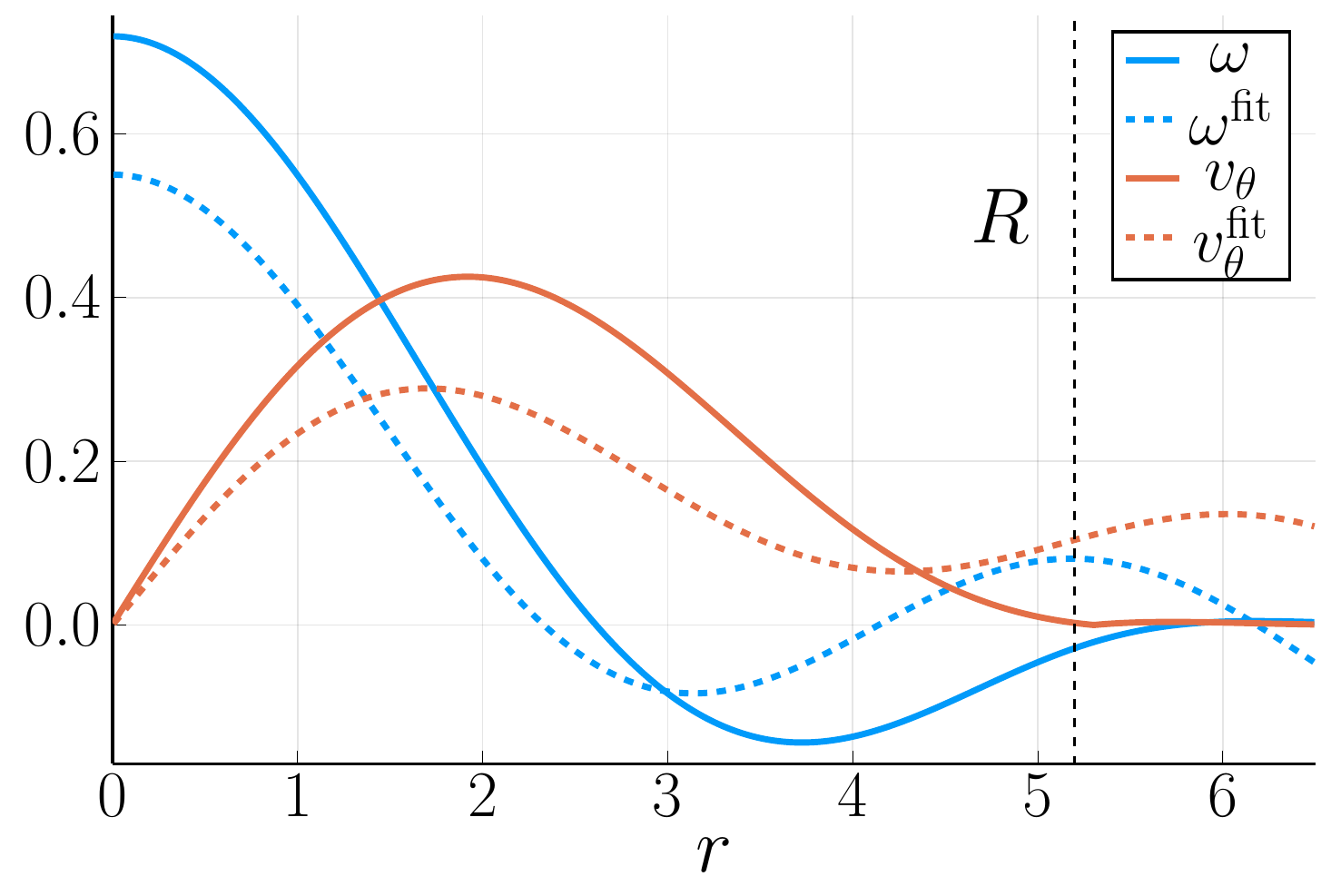}
    \end{minipage}
    \begin{minipage}[b]{0.48\linewidth}
        \centering
        \subcaption{
        } \label{fig:mode}
        \includegraphics[width = \hsize]{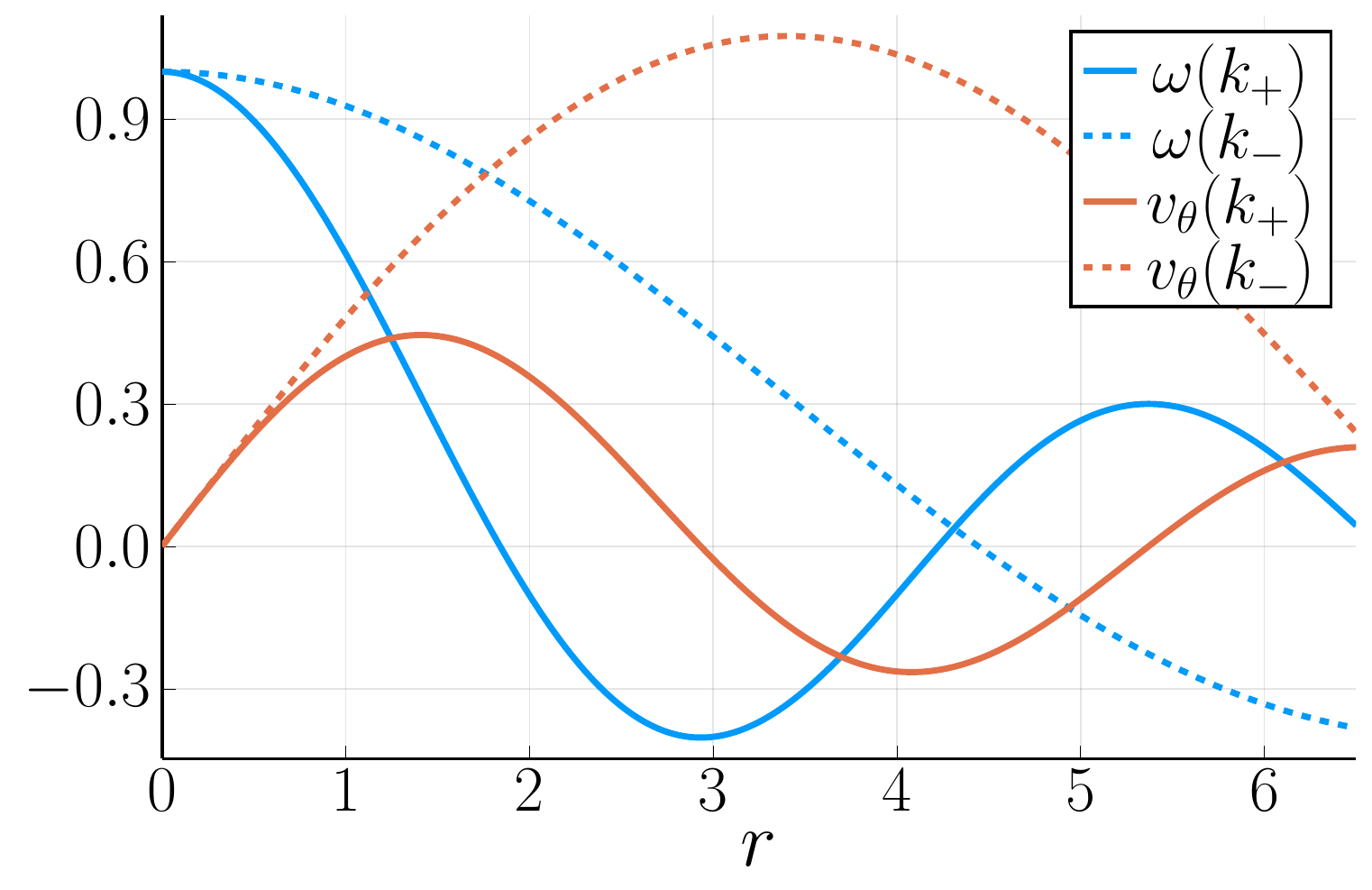}
    \end{minipage}
    \caption{
        Comparison between the single-vortex stationary state and analytic solutions for the linearized TTSH equation.
        \subref{fig:slice} Radial dependence of $v_\theta$ (blue) and $\omega$ (orange) for the single-vortex stationary state at $R = 5.2$ (solid) and those for the best-fit linear combination of the analytical solutions with $n=0$ (dashed). The fitting was carried out based on both $\vb{v}$ and $\omega$ normalized by their mean absolute values. The same set of $C_0^\pm$ is used for $v_\theta$ and $\omega$. \subref{fig:mode} Two independent modes of the analytical solutions, $v_{0,\theta}^\pm$ and $\omega_0^\pm$ for $k = k_{\pm}$.
    }
\end{figure}

It is useful to compare the velocity and vorticity fields in our single-vortex stationary state with those of an analytical solution to the linearized version of the TTSH equation \cite{reinken2020organizing}.
As described in \appref{sec:exactsolution}, the general stationary solution to the TTSH equation without nonlinear terms, expressed in terms of polar coordinates $(r,\ \theta)$, is given by
\begin{gather}
    \omega(\vb{x}) = \sum_{n=0}^\infty \sum_\pm C_n^\pm \omega_n^\pm(r,\ \theta), \\
    \omega_n^\pm(r,\ \theta) \equiv J_n\qty(k_\pm r) \cos{\Theta_n},
\end{gather}
and
\begin{gather}
    \vb{v}(\vb{x}) = \sum_{n=0}^\infty \sum_\pm C_n^\pm \vb{v}_n^\pm(r,\ \theta), \\
    v_{n,x}^\pm(r,\ \theta) = \frac{J'_n\qty(k_\pm r)}{k_\pm} \cos{\Theta_n} \sin{\theta} - \frac{nJ_n\qty(k_\pm r)}{k_\pm^2 r} \sin{\Theta_n} \cos{\theta}, \\
    v_{n,y}^\pm(r,\ \theta) = \frac{J'_n\qty(k_\pm r)}{k_\pm} \cos{\Theta_n} \cos{\theta} + \frac{nJ_n\qty(k_\pm r)}{k_\pm^2 r} \sin{\Theta_n} \sin{\theta},
\end{gather}
with $v_{n,x}^\pm$ and $v_{n,y}^\pm$ being the $x$ and $y$ components, respectively, of $\vb{v}_n^\pm$, $J_n$ the Bessel function of the first kind, $k_\pm \equiv \sqrt{1 \pm \sqrt{a}}$, and $\Theta_n \equiv n\theta + \mathrm{const}$.
Since the numerically observed single-vortex state is isotropic (i.e., independent of $\theta$), we are led to compare with the analytic solutions with $n=0$.
The result is displayed in \figref{fig:slice}.
Interestingly, we found that both the tangential component of the velocity field $v_\theta(r,\ \theta)$ and the vorticity field $\omega(r,\ \theta)$ have qualitative features in common with a linear combination of those analytic solutions, especially near the center, despite the existence of the nonlinear terms in our simulations.
This may partly be because the advection term $\lambda \vb{v} \vdot \grad{\omega}$ of \eqref{eq:mod2} vanishes for the isotropic solution.
However, since the analytic solutions cannot satisfy the boundary conditions for $\vb{v}$ and $\omega$ simultaneously, they cannot describe the numerical observation precisely.
We may argue that the existence of the cubic term $-b \vb{v}^2 \vb{v}$ of the TTSH equation may serve, effectively, for the Bessel-type solution to adjust itself to reconcile with the required boundary conditions.

\begin{figure*}[t!]
    \begin{minipage}[b]{0.58\linewidth}
        \centering
        \subcaption{} \label{fig:phase_zoom}
        \includegraphics[width = \hsize]{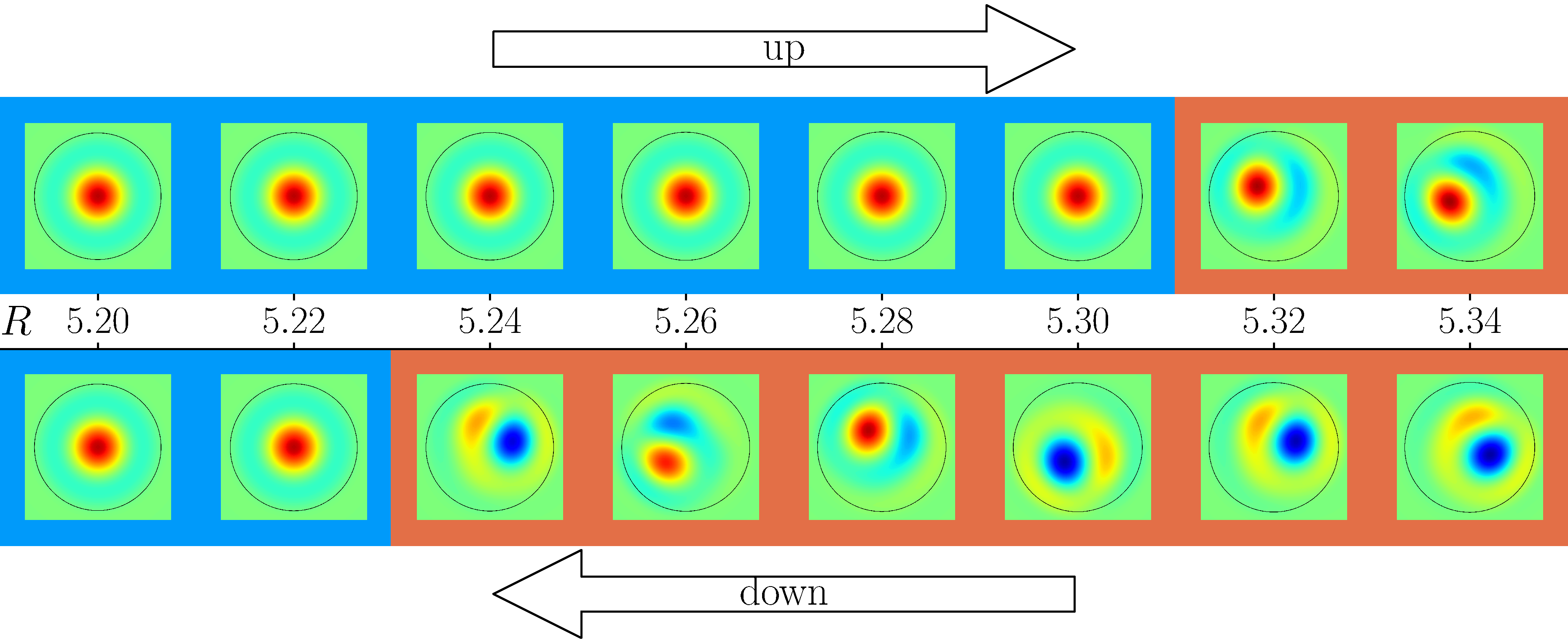}
    \end{minipage}
    \begin{minipage}[b]{0.38\linewidth}
        \centering
        \subcaption{
        } \label{fig:ord}
        \includegraphics[width = \hsize]{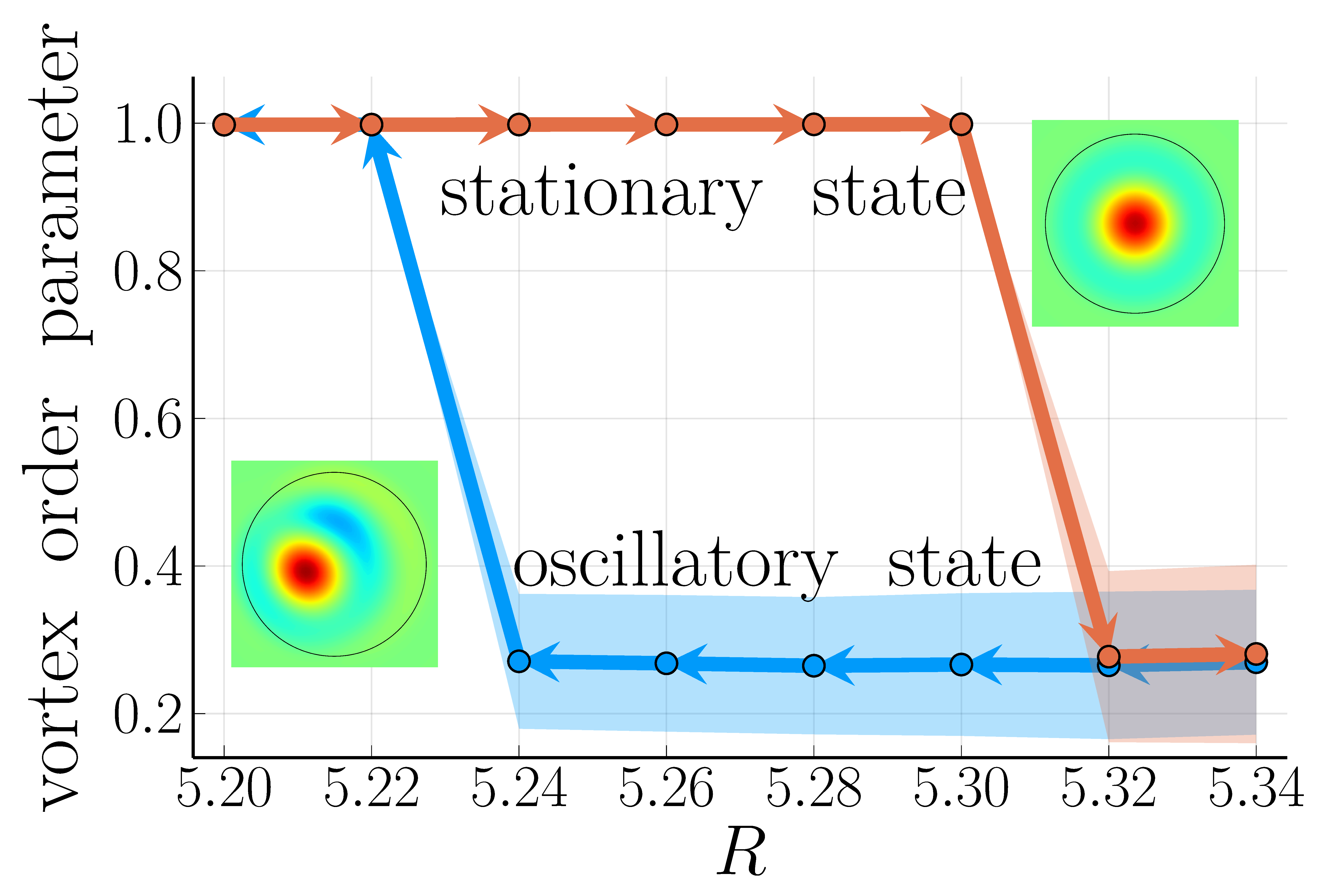}
    \end{minipage}\\
    \begin{minipage}[b]{0.24\linewidth}
        \centering
        \subcaption{
        } \label{fig:exp}
        \includegraphics[width = \hsize]{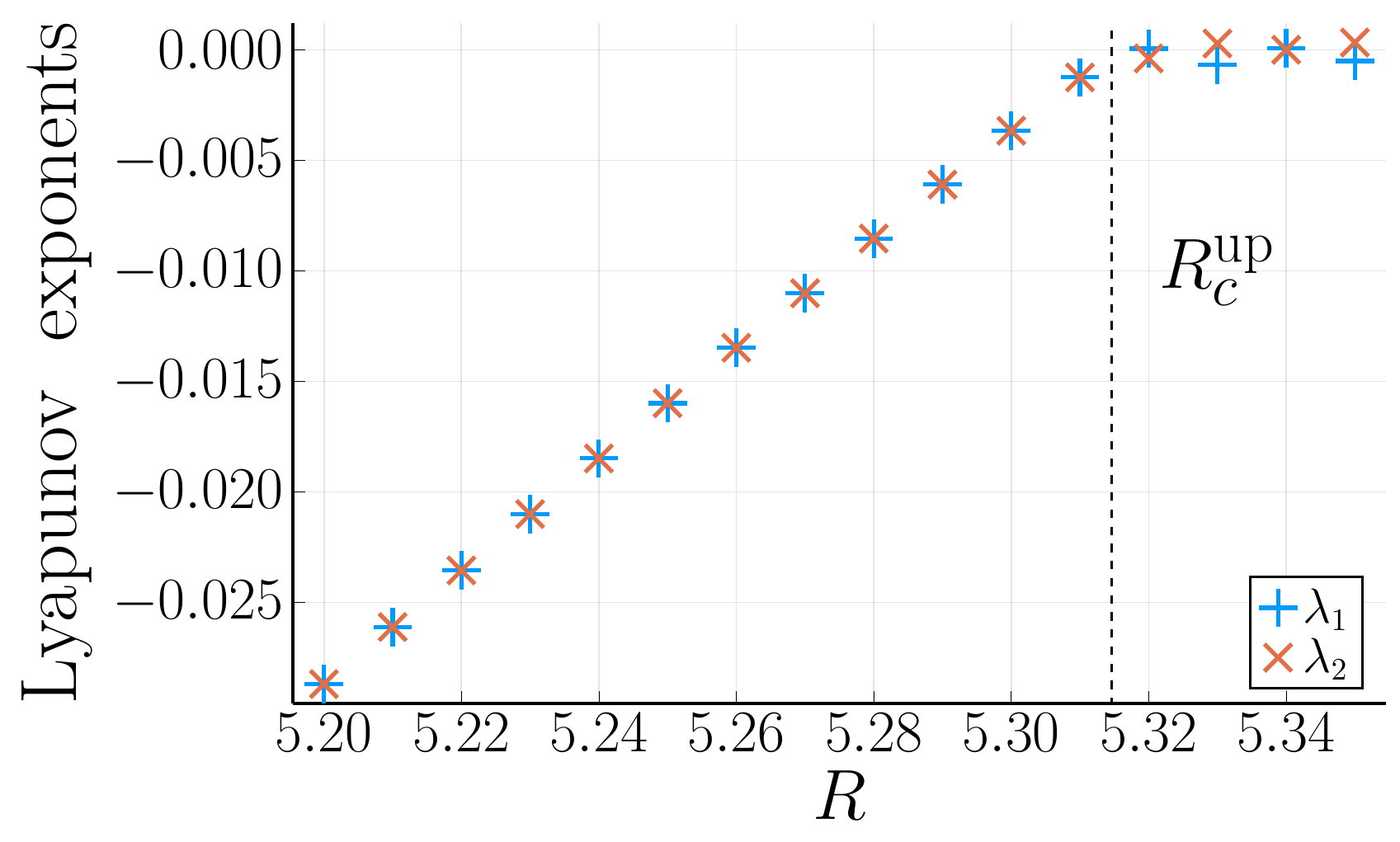}
    \end{minipage}
    \begin{minipage}[b]{0.24\linewidth}
        \centering
        \subcaption{
        } \label{fig:expu}
        \includegraphics[width = \hsize]{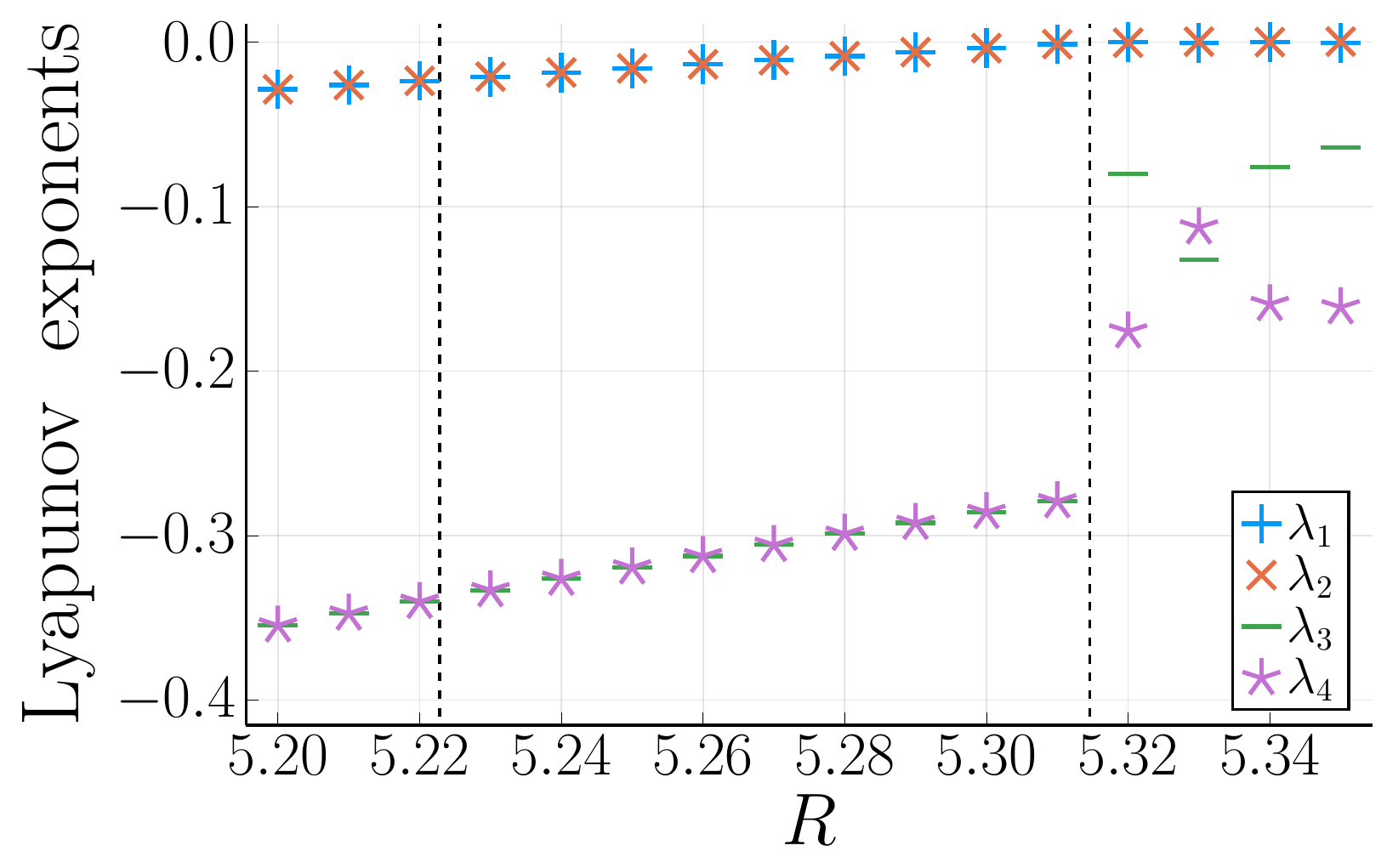}
    \end{minipage}
    \begin{minipage}[b]{0.24\linewidth}
        \centering
        \subcaption{
        } \label{fig:expd}
        \includegraphics[width = \hsize]{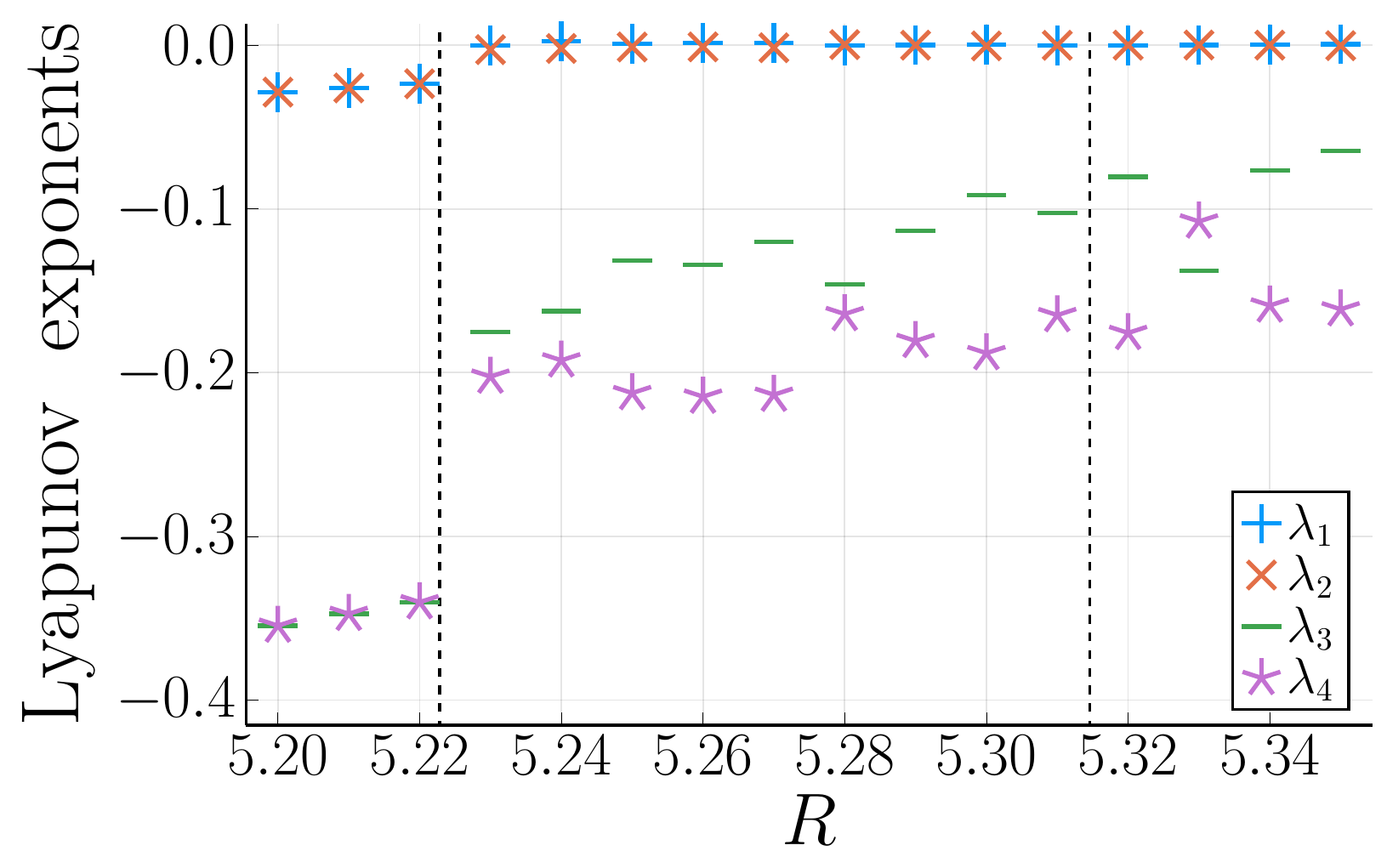}
    \end{minipage}
    \begin{minipage}[b]{0.24\linewidth}
        \centering
        \subcaption{
        } \label{fig:itv}
        \includegraphics[width = \hsize]{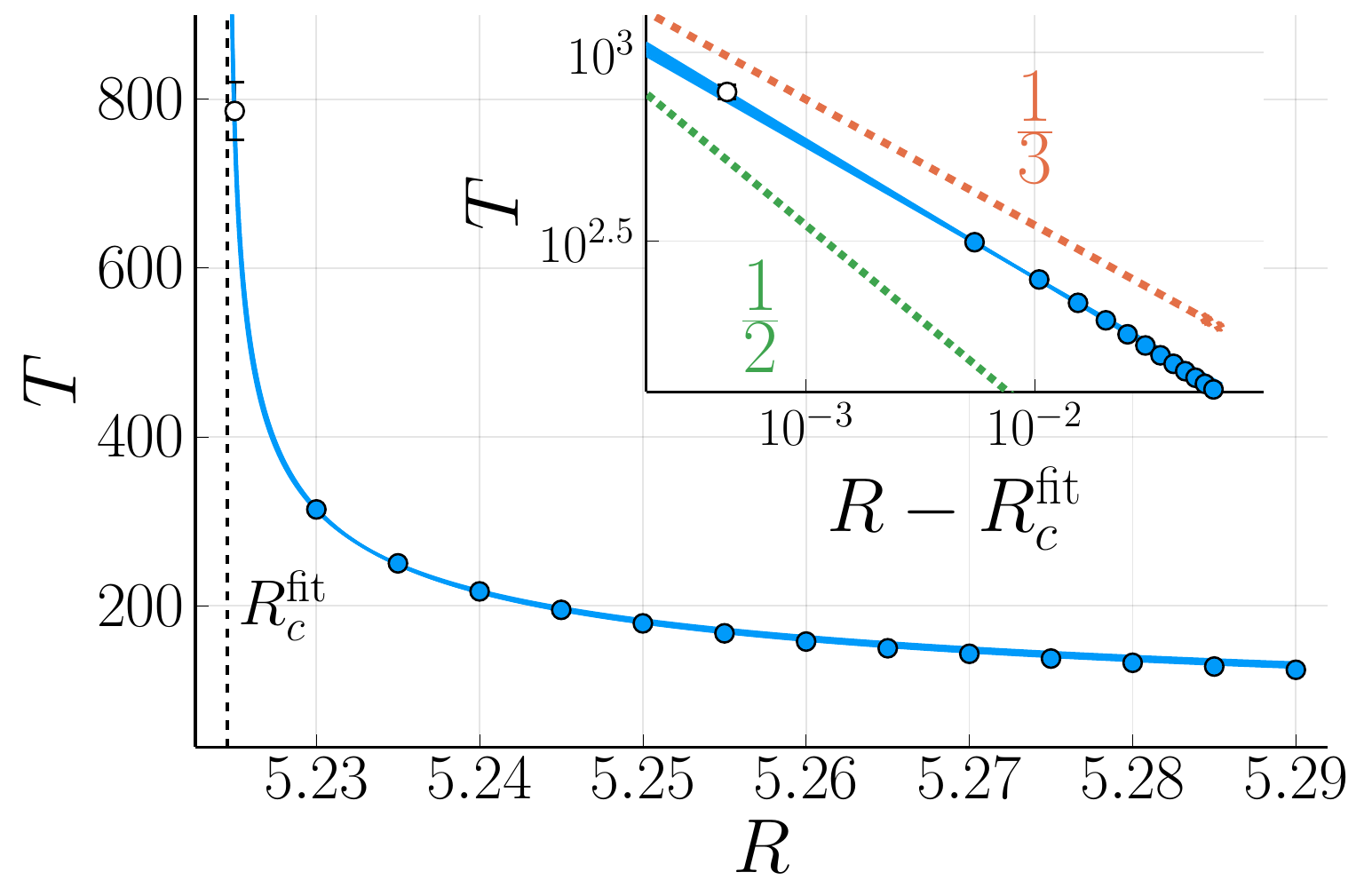}
    \end{minipage}
    \caption{
    Hysteretic transition between the single-vortex stationary state and the vortex-pair oscillatory state.
    \subref{fig:phase_zoom} Snapshots of the vorticity field (same color scale as \figref{fig:emoji_snap}) as $R$ is increased (top) or decreased (bottom).
    \subref{fig:ord} Vortex order parameter $\Psi$. The time-averaged values are shown. The shades indicate the temporal standard deviations.
    \subref{fig:exp}\subref{fig:expu} Lyapunov exponents for increasing $R$. The two and four largest exponents are shown in \subref{fig:exp} and \subref{fig:expu}, respectively. The first two exponents increase linearly with $R$ and cross zero at $R = R_c^{\mathrm{up}}$, consistently with the case of the subcritical Hopf bifurcation.
    \subref{fig:expd} Four largest Lyapunov exponents for decreasing $R$.
    \subref{fig:itv} Oscillation period $T$ above the lower transition point $R_c^{\mathrm{down}}$. Its standard deviation is indicated by error bars, visible only for $R=5.225$ (open symbol) where the oscillation was fluctuating.
    Blue solid lines indicate results of fitting by $T \propto \qty|R - R_c^{\mathrm{fit}}|^{-p}$, for several choices of data points to use (hence many blue lines are drawn).
    Inset: The same data in the log-log scale. Two dashed lines are guides for the eyes corresponding to power laws with exponent $1/2$ (green) and $1/3$ (orange).}
\end{figure*}

\subsection{transition to vortex-pair oscillatory state}  \label{sec:hysteresis}

As $R$ is increased from the single-vortex stationary state, the system undergoes a transition to the vortex-pair oscillatory state (orange box in \figref{fig:phase_anime}).
This state consists of two oscillating vortices with vorticities of different signs, as already shown in \figrefs{fig:anime} and \ref{fig:ring}, as well as in \movrefs{S2 and S4} \cite{SM}.

The transition between the single-vortex stationary state and the vortex-pair oscillatory state turned out to be hysteretic, as illustrated in \figref{fig:phase_zoom}.
A useful quantity to capture this transition is the vortex order parameter \cite{wioland2013confinement}
\begin{equation}
    \Psi = \frac{1}{1 - \frac{2}{\pi}}\qty(\frac{\sum_{\vb{x}} \qty|\vb{e}_\theta(\vb{x}) \vdot \vb{v} \qty(\vb{x})|}{\sum_{\vb{x}} \qty|\vb{v} \qty(\vb{x})|} - \frac{2}{\pi}),
\end{equation}
where $\vb{e}_\theta(\vb{x})$ is the azimuthal unit vector and the sum is taken over the entire space.
By construction, $\Psi = 1$ if the velocity field is completely azimuthal, whereas $\Psi = 0$ if it is completely disordered.
To investigate the hysteresis, we ran simulations sequentially, as follows.
To go up, starting from the steady state at $R = 5.20$, we increased $R$ by $\Delta R = 0.02$ and measured $\Psi$ typically over 5-10 periods after the system reached the steady state.
We repeated this step until $R=5.34$.
To go down, we did the same in the opposite direction.
The result is shown in \figref{fig:ord}, which clearly shows the hysteresis.
From this, we approximately estimated the lower and upper transition points at $R_c^{\mathrm{down}} \simeq 5.23$ and $R_c^{\mathrm{up}} \simeq 5.31$, respectively.

To quantitatively characterize the dynamic aspect of this transition, we measured the Lyapunov exponents, i.e., the exponential growth rates of infinitesimal perturbations to the solution, which can be a direct clue to determine the type of bifurcation underlying this transition \cite{Strogatz-Book2001}. To calculate them, following the standard method \cite{shimada1979numerical,Benettin.etal-M1980}, we simulated the time evolution of independent perturbations $\delta\omega_i(\vb{x},\ t)$ ($i$: index) along with $\omega(\vb{x},\ t)$ and measured the exponential growth rates using the QR decomposition. The Lyapunov exponents $\lambda_i$ were then obtained in ascending order. This procedure is even more costly than the main calculation and took days even though we had access to cutting-edge GPUs.

The result is shown in \figref{fig:exp} for increasing $R$, with $\Delta R=0.01$.
This shows that the two largest exponents $\lambda_1,\ \lambda_2$, which are negative for $R < R_c^{\mathrm{up}}$ as expected, increase linearly with $R$ and reach zero at $R = R_c^{\mathrm{up}}$. This behavior is consistent with the subcritical Hopf bifurcation, which indeed shows a hysteretic transition to an oscillatory state \cite{Strogatz-Book2001}.
It is also notable that the oscillatory state involves \textit{two} vanishing exponents, despite the absence of quasiperiodic behavior.
We consider that the rotational symmetry of the system introduces the second vanishing exponent, in addition to the one corresponding to the time translation symmetry.
Regarding the third and fourth exponents  $\lambda_3,\ \lambda_4$, they also increase linearly with $R$ up to $R_c^{\mathrm{up}}$ [\figref{fig:expu}].
Interestingly, extrapolation of this linear dependence suggests that it would cross zero at $R \simeq 5.65$, which is close to the transition point to the state with four oscillating vortices.

\begin{figure*}[tb]
    \centering
    \includegraphics[width = \hsize]{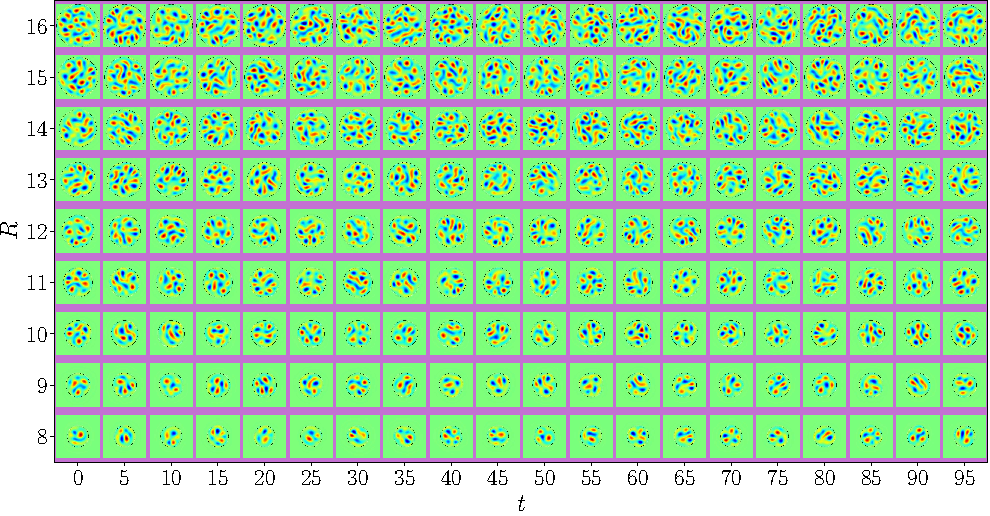}
    \caption{
        Time series of the vorticity field for $R \ge 8$ (same color scale as \figref{fig:emoji_snap}).
        Note that the results presented here are affected by relatively strong effects of the periodic boundary, because $R$ is close to $L / 2$ and the vorticity field does not decay sufficiently at the periodic boundary.
    } \label{fig:anime_big}
\end{figure*}

For decreasing $R$, the result is shown in \figref{fig:expd}. Given that the transition at $R_c^{\mathrm{up}}$ was consistent with the subcritical Hopf bifurcation, the standard scenario from the low-dimensional dynamical systems theory suggests that the lower transition may be described by the saddle-node bifurcation, in which case the first negative exponent increases as $\propto \qty|R - R_c^{\mathrm{down}}|^{1/2}$ when approaching $R_c^{\mathrm{down}}$ from above \cite{Strogatz-Book2001}. However, it was not our case: The first negative exponent $\lambda_3$ does not approach zero but remains at $\simeq -0.2$ as displayed in \figref{fig:expd}.
We also measured the oscillation period $T$ as a function of $R$ [\figref{fig:itv}], using the half smaller discretization interval $\Delta x$ to reduce discretization effect (with the edge length $N \Delta x$ kept unchanged; see \tblref{tab:param} in \appref{sec:notation}).
If the transition were the saddle-node bifurcation, we would expect $T \propto \qty|R - R_c^{\mathrm{down}}|^{-1/2}$.
However, while our data indeed show seemingly power-law divergence near $R_c^{\mathrm{down}}$, i.e., $T \propto \qty|R - R_c^{\mathrm{fit}}|^{-p}$, we estimated the exponent $p$ at $p = 0.35(1)$.
This was obtained by varying the range of fitting, on which the result hardly depends, except that we obtained $p=0.39(1)$ if the point closest to transition, at which the oscillation seemed less stable, was excluded.
This suggests that the bifurcation we observed may not be understood within the framework of the low-dimensional dynamical systems theory. In other words, this hints at a hitherto unknown bifurcation in high-dimensional dynamical systems.

\subsection{route to turbulence} \label{sec:route}

Finally we comment on the route to turbulence observed in this circular confinement.
As already described in \secref{sec:overview} and \figref{fig:phase_anime}, the first nontrivial state observed in this geometry is the single-vortex stationary state ($R = 5.2$).
As $R$ is increased, the system first undergoes a hysteretic transition to the vortex-pair oscillatory state ($5.23 \lesssim R \lesssim 5.31$).
This state is periodic for $R$ close to the transition, while chaotic modulation may be added for larger $R$.
From $R=5.8$, the system has four vortices and shows multiple transitions among different dynamical states (periodic, quasiperiodic, and chaotic states) in a reentrant manner.
The number of vortices increases further from $R=7.6$ and the system now stays in the chaotic state (\figref{fig:anime_big}).
This state continuously shifts to active turbulence in the bulk limit ($R \to \infty$).

When compared to the route to turbulence for the Navier-Stokes turbulence, while the emergence of periodic and quasiperiodic states is also seen in the Ruelle-Takens-Newhouse scenario \cite{ruelle1971nature,Eckmann-RMP1981}, the rest of the observations do not correspond to any well-known scenario.
Instead, we note that similar reentrant transitions among periodic, quasiperiodic, and chaotic states were observed in the Kuramoto-Sivashinsky equation \cite{Hyman.Nicolaenko-PD1986}, as well as in a numerical simulation of the Navier-Stokes equation when a so-called high-symmetry condition was imposed on the flow field \cite{kida1989route}. Although we did not impose such a condition, our highly symmetric circular geometry may be relevant to the reentrant behavior that characterizes the observed route to turbulence.

\begin{figure}[tb]
    \centering
    \begin{minipage}[b]{0.48\linewidth}
        \centering
        \subcaption{} \label{fig:pdef}
        \includegraphics[width = \hsize]{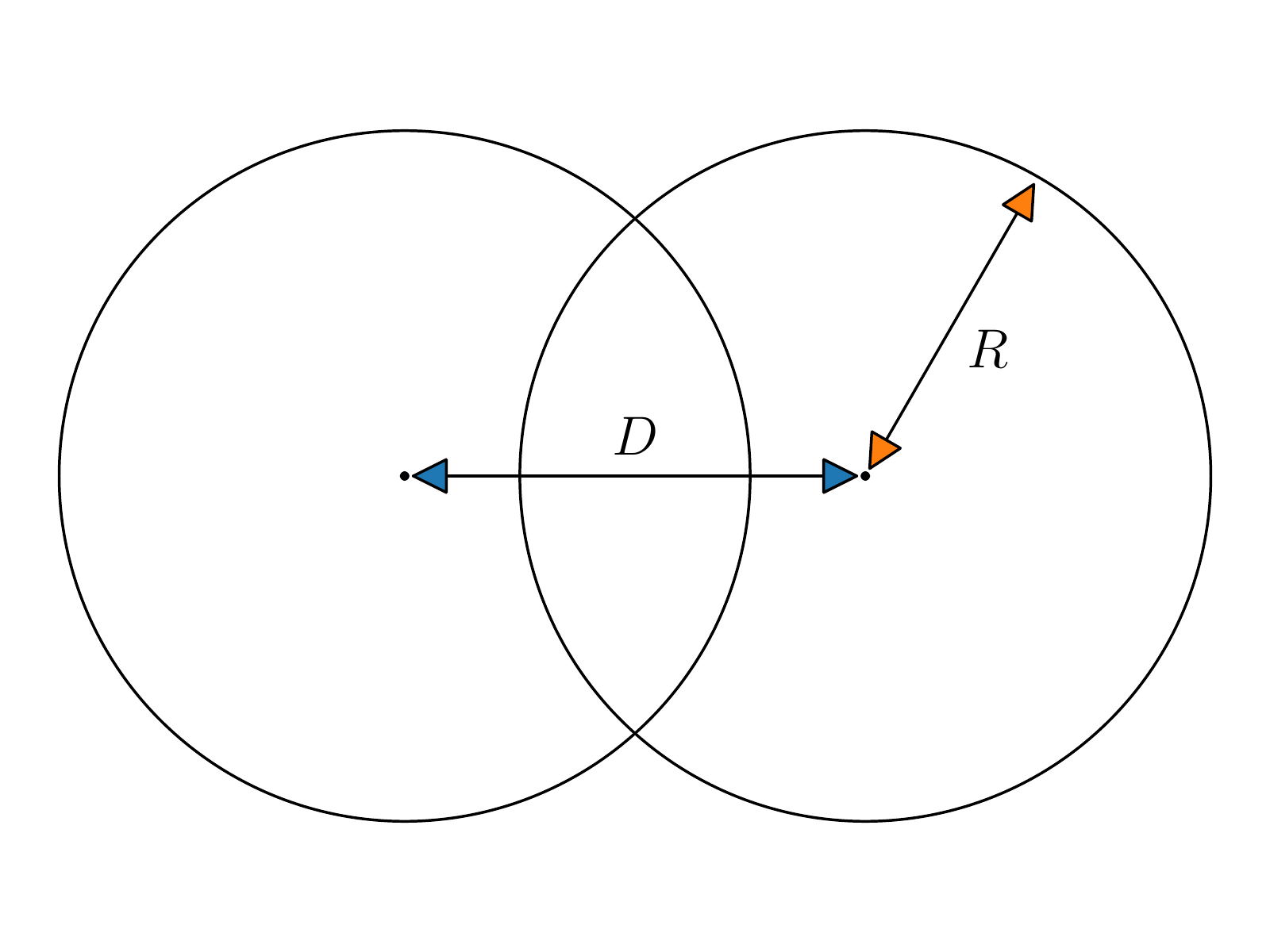}
    \end{minipage}
    \begin{minipage}[b]{0.48\linewidth}
        \centering
        \subcaption{} \label{fig:soenord}
        \includegraphics[width = \hsize]{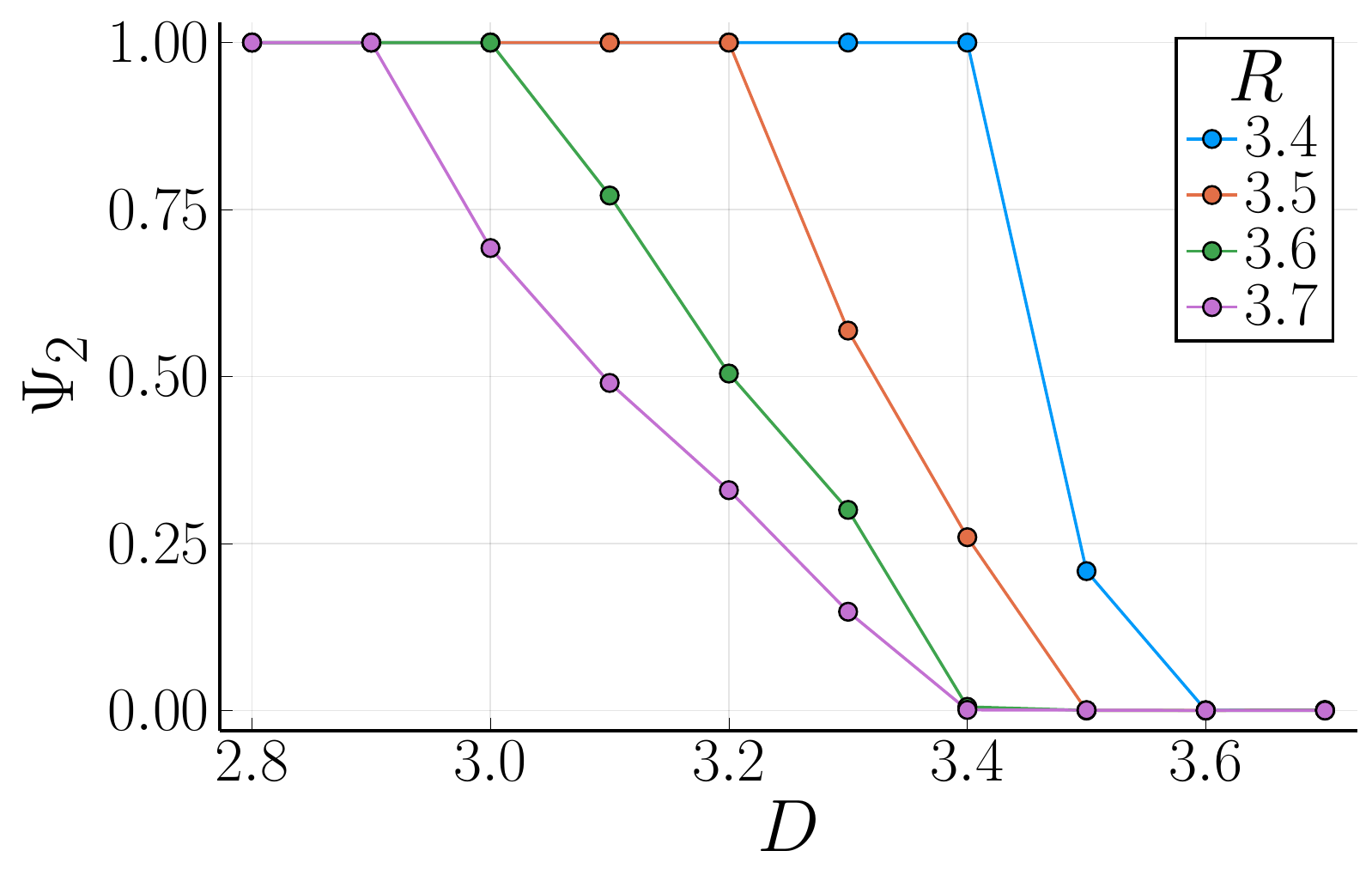}
    \end{minipage}
    \begin{minipage}[b]{\linewidth}
        \centering
        \subcaption{} \label{fig:soen}
        \includegraphics[width = \hsize]{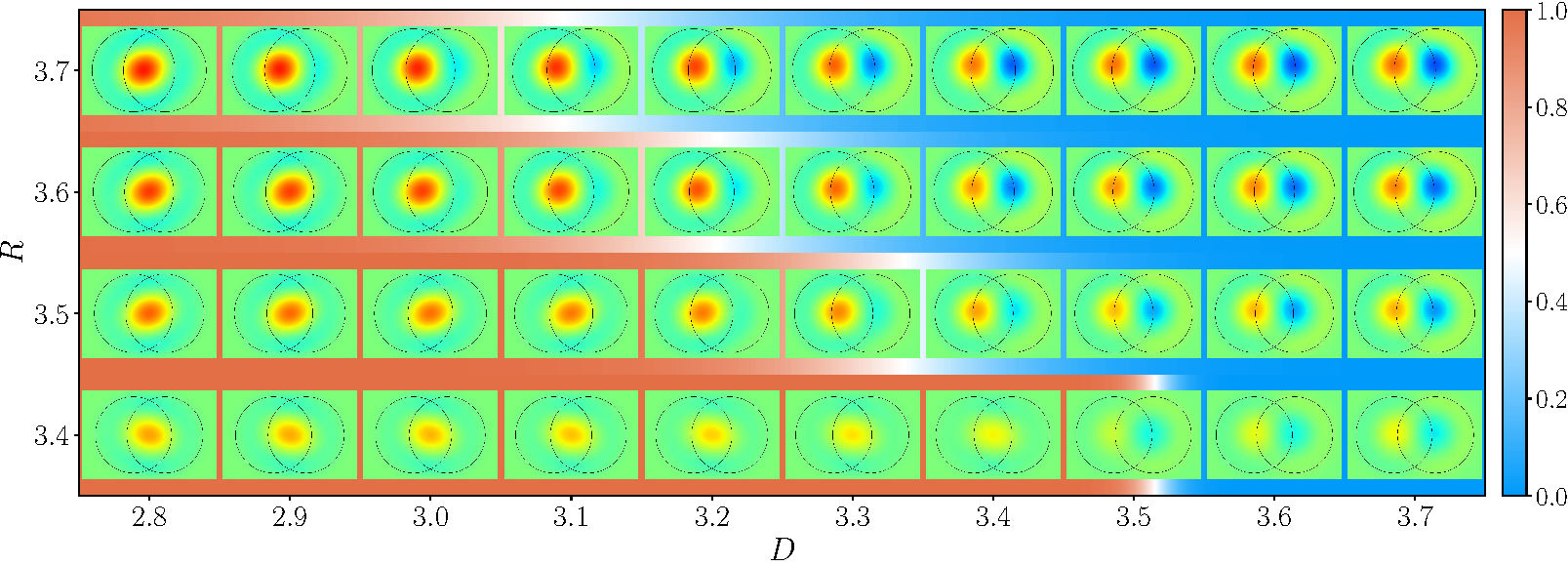}
    \end{minipage}
    \caption{
        Confinement in dumbbell.
        \subref{fig:pdef} Outline of the dumbbell-shaped boundary made of two overlapping circles and the associated parameters.
        \subref{fig:soenord} Order parameter $\Psi_2$ vs the center-to-center distance $D$ for different radii $R$.
        By construction, $\Psi_2 = 1$ for the ferromagnetic flow and $\Psi_2 = 0$ for the antiferromagnetic one.
        \subref{fig:soen} Typical vorticity snapshots taken after relaxation to the stationary state, for different radii $R$ and center-to-center distances $D$ (same color scale as \figref{fig:emoji_snap}). The background color indicates the value of $\Psi_2$ according to the color bar.
        For simplicity, some snapshots are displayed after mirror reflection so that the vorticity is always positive in the left half and that the stronger vortex has the positive vorticity (the latter can be realized by applying $y \mapsto -y$).}
\end{figure}

\section{confinement in dumbbell}  \label{sec:dumbbell}

\subsection{motivations and background}
Here we turn our eyes to a dumbbell-shaped confinement as shown in \figref{fig:pdef}, which consists of two overlapping circles of radius $R$ with the centers separated by distance $D$. This test case serves as a benchmark to demonstrate the ability of our method to implement a complex geometry without handcrafting the mask $K(\vb{x})$.

In the case of bacterial turbulence enclosed in a PDMS device \cite{beppu2017geometry,beppu2021edge}, it was reported that a pair of vortices was formed in dumbbell-shaped boundaries, each vorticity peak located near the center of the circle.
With a threshold distance $D_c\qty(R) = \sqrt{2} R$, the signs of the two vorticity peaks were identical (ferromagnetic vortex order) for $D < D_c$ and opposite (antiferromagnetic vortex order) for $D > D_c$ \cite{beppu2017geometry}.
It is therefore of interest to see if a similar transition takes place in our hydrodynamic setup.

Our simulations on the dumbbell confinement were performed as follows.
Similarly to the bacterial experiments \cite{beppu2017geometry,beppu2021edge}, we set the cavity radius near the characteristic size of a single vortex, $R \approx \pi$.
Note that, although even a single vortex could not appear at such small radii $R$ in the case of the circular confinement (see \secref{sec:overview}), in the dumbbell confinement we do observe vortices, presumably because of the relatively larger total area of the confinement.
In the simulations reported below, we started from a random initial state and discarded transients to ensure that the system is in a steady state, for each choice of $R$ and $D$.

\subsection{results}
Figure~\ref{fig:soen} displays our simulation results.
At each fixed $R$, for small $D$ we observed a single vortex formed near the center of the dumbbell, and for large $D$ a pair of vortices is formed near the centers of the two cavities.
In either case, the vortices are stationary in the steady state.
The latter, vortex-pair state corresponds to the antiferromagnetic state observed in previous experiments \cite{beppu2017geometry,beppu2021edge}.
Regarding the former, single-vortex state, it is analogous to the ferromagnetic state in the sense that the vorticity has the same sign in the entire region of the dumbbell, but we did not observe a split of vortices as in the experiments.
This difference may be attributed to the presence of the edge flow under the conditions of Beppu \etals experiments \cite{beppu2017geometry,beppu2021edge}, while our choice of the boundary conditions is based on another experimental setup by Nishiguchi and coworkers \cite{nishiguchi2018engineering,reinken2020organizing}.

A difference is also seen in the transition between the two states.
In our simulations, as shown in \figref{fig:soen}, the threshold distance $D_c(R)$ decreases with increasing $R$, while it was increasing as $D_c(R) = \sqrt{2}R$ in Beppu \etals experiments \cite{beppu2017geometry}.
To quantify this observation, we measured an order parameter specialized for this case, defined by $\Psi_2 = \left|\omega(\vb{x}^{\ast}_{\mathrm{left}}) + \omega(\vb{x}^{\ast}_{\mathrm{right}})\right| / \qty(|\omega|(\vb{x}^{\ast}_{\mathrm{left}}) + |\omega|(\vb{x}^{\ast}_{\mathrm{right}}))$ with $\vb{x}^{\ast}_{\mathrm{left/right}} = \mathrm{argmax}_{\mathrm{left/right}} |\omega|$.
Here, the subscript ``left'' and ``right'' stands for the area within $D / 2$ from the center of either cavity.
By construction, $\Psi_2 = 1$ for the ferromagnetic flow and $\Psi_2 = 0$ for the antiferromagnetic one.
The results in \figref{fig:soenord} clearly demonstrate that $\Psi_2$ transitions from $\approx 1$ to $\approx 0$ as $D$ is increased, with $D_c(R)$ decreasing with increasing $R$.
Therefore, our $D_c(R)$ is ruled by a law different from the scaling $D_c(R) \propto R$ observed in Beppu \etals experiments \cite{beppu2017geometry,beppu2021edge}.
Beppu \etal \cite{beppu2017geometry} accounted for the relation $D_c(R) = \sqrt{2}R$ on the basis of edge currents, i.e., tangential flow on the boundary, which is incompatible with the boundary conditions chosen here (see discussions in \secref{sec:edge}).
It is therefore reasonable to consider that the absence of edge current in our setup may be responsible for this difference.

\section{discussions} \label{sec:discussion}

\subsection{edge current} \label{sec:edge}

Here we discuss the edge current, i.e., tangential flow along the boundary, typically seen in experiments under circular confinements \cite{wioland2013confinement,enkeleida2014fluid,beppu2017geometry,beppu2021edge}.
The edge current involves non-vanishing tangential velocity at the boundary, $v_\theta \neq 0$ in the case of the circular confinement.
In the aforementioned experiments, bacteria actually swim along the boundary.
The formation of counter-rotating double layers has also been reported in droplet suspensions of {\it B. subtilis} \cite{wioland2013confinement}, while no such counter-rotating layers were reported in the case of {\it E. coli} confined in a microfluidic device \cite{beppu2017geometry,beppu2021edge}.
In contrast, in our simulations, $v_\theta$ continuously decays to zero due to the damping terms, without edge current nor counter-rotating layer.
This difference clearly results from our choice of the boundary conditions, $\vb{v} = \vb{0}$ and $\omega = 0$, which were deduced from the experimental observation of dense {\it B. subtilis} suspensions placed on a substrate with microfabricated pillars and bordered by a liquid-air interface \cite{nishiguchi2018engineering,reinken2020organizing}.
Although counter-rotating layers were not observed in the low-magnification microscopy carried out in these experiments, the possible existence of such layers may have led to our boundary conditions, as discussed in \citref{reinken2020organizing}. Therefore, we expect that our numerical simulations provide predictions for experimental conditions similar to those in \citrefs{nishiguchi2018engineering,reinken2020organizing}, rather than those in the existing experiments of circularly confined bacteria \cite{wioland2013confinement,enkeleida2014fluid,beppu2017geometry,beppu2021edge}.

To extend the model to deal with the case with edge currents, we may (1) introduce a slip velocity to allow a nonvanishing $v_\theta$ at $r = R$ or (2) use another governing equation that does not suppress high-wavenumber variations, thus allowing the existence of current near the no-slip boundary.
We tested the approach (1) by using a damping scheme that only removes the radial velocity component $v_r$.
However, we were unable to carry out physically sound simulations in this case, because the tangential component $v_\theta$ penetrated deep inside the damped zone.
In passing, we note that this boundary condition allowed us to set the circle radius $R$ smaller than the minimum value reported in \secref{sec:ConfinementInCircle} to generate a vortex.
This may be related to the fact that, for the no-slip condition, the identity $\int_{r < R} \omega dxdy = \oint_{r = R} \vb{v} \vdot d\vb{x} = 0$ guarantees that positive and negative vorticities must exist in the same amount inside the confinement, while for the slip case $\omega$ can escape from the confinement despite the damping.
Regarding the approach (2), we may remove the $\nabla^4 \vb{v}$ term of the TTSH equation.
The equation then reduces to the incompressible version of the Toner-Tu equation, for which vortices were reported to appear in bulk systems \cite{besse2022metastability} unlike the original, compressible Toner-Tu equation \cite{toner1995longrange,toner1998flocks,toner2012reanalysis}.
These approaches may deepen our understanding of effect of edge current on structures and dynamics of active fluids in confined geometries.

\subsection{implementation of the boundary conditions}

One may wonder if the boundary conditions, whether slip or no-slip, can be imposed more directly, without resorting to the damping scheme.
In this case, we have to deal with arbitrary-shaped boundaries directly.
In the literature, such a method has been pursued in broader contexts.
For example, in computational fluid dynamics  \cite{versteeg2007introduction}, an arbitrary-shaped boundary is typically realized by a tailored mesh, and the model is discretized and integrated on it.
For the TTSH equation, however, the discretization of the fourth-order derivative and the boundary condition of vanishing vorticity is not straightforward on such a mesh.
More specifically, one may want to use the second-order derivative of $\vb{v}$ (to use the computationally efficient Laplacian) on the boundary, but this is incompatible with the discrete fourth-order derivative of the time evolution equation. More sophisticated algorithms may be devised, but we suspect that it is difficult to avoid uncontrollable approximations and that such algorithms are more costly than the pseudospectral method we adopted.

\section{concluding remarks}  \label{sec:conclusion}

In this work, we realized numerical simulations of the TTSH equation with arbitrary-shaped boundaries, and presented results for the two representative test cases, namely the circular and dumbbell confinements.

From the computational perspective, first we emphasize that GPU implementation makes calculations fast and affordable, without the need to use supercomputers or to wait for weeks. Indeed, this paper contains several simulations that would take months or even years using workstation-class CPU (typically $10^1 \sim 10^2$ threads, $3$~GHz).

Let us conclude this paper by discussing physical implications of the work.
Our numerical investigation with the TTSH equation has succeeded in reproducing the emergent vortex order reported in the experiments, at least qualitatively. At the same time it has highlighted that the slight difference in the confinement geometry can give rise to quantitatively different vortex structures.
Specifically, we have explored a novel route to chaos and turbulence under the circular confinement in the TTSH equation.
This route starts with a hysteretic transition from an ordered vortex to an oscillating pair of vortices, consistent with the subcritical Hopf bifurcation.
This is followed by reentrant transitions across periodic, quasiperiodic, and chaotic oscillations, until the system finally reaches the active turbulent state.
Since our boundary conditions are the ones inferred from the specific experimental realization reported in \citref{reinken2020organizing}, it is an important future task to investigate how robust our findings are, for other models and boundary conditions that may be more suitable for other experimental setups.
In this context, it is interesting to note that a Hopf bifurcation was also reported for another polar active fluid model without confinement but variable activity \cite{giomi2012polar}, suggesting some extent of universality in the route to polar active turbulence via oscillatory states.
We anticipate that our results will contribute to the fundamental understanding of how turbulent structure develops in active matter systems.

\appendix

\section{analytic solution for the linearized TTSH equation}
\label{sec:exactsolution}

Following \citref{reinken2020organizing}, here we describe analytic solutions for the linearized TTSH equation, which amounts to setting $\qty(b,\ \lambda) = \qty(0,\ 0)$ in \eqref{eq:orig2}.
Note that the dropped nonlinearity cannot be regarded as a perturbation.
Therefore, the analytic solutions described in this section may not necessarily represent the numerical observations even approximately. Nevertheless, these analytic solutions help interpret some numerical results presented in \secref{sec:circle}.

The linearized TTSH equation for the vorticity, with the stationarity condition, reads:
\begin{equation}
    0 = a \omega - \qty(1 + \laplacian)^2 \omega. \label{eq:lin2}
\end{equation}
This can be easily solved in the polar coordinate system $\qty(r,\ \theta)$, on the basis of the real-valued eigenfunctions of the Laplacian, $J_n \qty(kr) \cos\qty(n\theta + \mathrm{const.})$, where $J_n$ is the Bessel function of the first kind, $k>0$, and $n = 0,\ 1,\ \dots$.
The corresponding eigenvalue is $-k^2$.
Therefore, a solution to \eqref{eq:lin2} needs to satisfy $0 = a - (1 - k^2)^2$, i.e., $k = k_\pm \equiv \sqrt{1 \pm \sqrt{a}}$.
The general solution is then given by linear combinations of them:
\begin{gather}
    \omega(\vb{x}) = \sum_{n=0}^\infty \sum_\pm C_n^\pm \omega_n^\pm(r,\ \theta), \label{eq:ex0} \\
    \omega_n^\pm(r,\ \theta) \equiv J_n\qty(k_\pm r) \cos{\Theta_n}, \label{eq:ex1}
\end{gather}
with $\Theta_n \equiv n\theta + \mathrm{const}.$

Moreover, in the case of the two-dimensional incompressible fluid studied in this work, the velocity field $\vb{v}(\vb{x})$ can be reconstructed from the vorticity field $\omega(\vb{x})$ by using the stream function.
First we obtain the stream function $\psi(\vb{x})$ by solving $\laplacian{\psi} = -\omega$, then it follows that $v_x = \partial_y \psi$ and $v_y = -\partial_x \psi$.
This gives
\begin{gather}
    \vb{v}(\vb{x}) = \sum_{n=0}^\infty \sum_\pm C_n^\pm \vb{v}_n^\pm(r,\ \theta), \\
    v_{n,x}^\pm(r,\ \theta) = \frac{J'_n\qty(k_\pm r)}{k_\pm} \cos{\Theta_n} \sin{\theta} - \frac{nJ_n\qty(k_\pm r)}{k_\pm^2 r} \sin{\Theta_n} \cos{\theta}, \\
    v_{n,y}^\pm(r,\ \theta) = \frac{J'_n\qty(k_\pm r)}{k_\pm} \cos{\Theta_n} \cos{\theta} + \frac{nJ_n\qty(k_\pm r)}{k_\pm^2 r} \sin{\Theta_n} \sin{\theta},
\end{gather}
using the same $C_n^\pm$ as \eqref{eq:ex0}.
Here, $v_{n,x}^\pm$ and $v_{n,y}^\pm$ are the $x$ and $y$ components, respectively, of $\vb{v}_n^\pm$.

In relation to our results for the circular domain with finite radius $R$ (\secref{sec:circle}), we first note that, if any special solutions $\omega_n^\pm(r,\ \theta)$ and $\vb{v}_n^\pm(r,\ \theta)$ satisfy $\omega_n^\pm(R,\ \theta) = 0$ and $\vb{v}_n^\pm(R,\ \theta)=0$ simultaneously, this can be regarded as a solution for the circular domain case, aside from the ignored nonlinearity.
However, these two conditions are actually never satisfied simultaneously, because for all $x>0$ the roots of $J_n\qty(x)$ and $J'_n\qty(x)$ do not coincide.
We may make $\omega(\vb{x})$ and $\vb{v}(\vb{x})$ closer to zero at $r=R$ by combining the two modes with $k = k_\pm$, but the boundary conditions cannot be exactly satisfied.
This point is commented in \secref{sec:phaseI}.

\section{notation}  \label{sec:notation}
\tblref{tab:param} summarizes the parameters and symbols used in the paper.

\begin{table}[H]
    \caption{Table of parameters and symbols.}
    \label{tab:param}
    \begin{tabular}{ll}\hline
        symbol                                                            & description                               \\ \hline
        $\lambda = 9$                                                     & advection strength                        \\
        $a = 0.5$                                                         & activity parameter                        \\
        $b = 1.6$                                                         & nonlinearity parameter                    \\
        $\gamma_{\vb{v}} = 40$                                            & velocity damping strength                 \\
        $\gamma_{\omega} = 4$                                             & vorticity damping strength                \\
        $N = 8192^\mathrm{a}$                                             & number of grid points per dimension       \\
        $\Delta x = 0.005^\mathrm{a}$                                     & spatial discretization interval           \\
        $\Delta t = 0.01$                                                 & temporal discretization interval          \\
        $L = N \Delta x$                                                  & edge length of the calculation area       \\
        $\hat{*},\ \dft{*}$                                               & discrete Fourier transform of $*$         \\
        $\idft{*}$                                                        & inverse discrete Fourier transform of $*$ \\
        $\vb{k} \in \qty[-\frac{\pi}{\Delta x},\ \frac{\pi}{\Delta x}]^2$ & DFT wavenumber                            \\
        $\Delta k = \frac{2\pi}{L}$                                       & discretization interval of $\vb{k}$       \\
        $K(\vb{x})$                                                       & preprocessed damping mask                 \\ \hline
    \end{tabular}\\
    $^\mathrm{a}$ $(N,\ \Delta x) = (16384,\ 0.0025)$ were used for the oscillation period $T$ shown in \figref{fig:itv}.\\
\end{table}

\newpage

\section{calculation scheme} \label{sec:scheme}
\algref{alg:int} is our integration scheme for the TTSH equation based on the pseudospectral method and the Euler method.

\begin{algorithm}[H]
    \begin{algorithmic}[1]
        \REQUIRE{$\qty(\vb{v} \qty(t),\ \hat{\omega} \qty(t))$}
        \STATE $\vb{rhs}_1 = -b \vb{v}^2 \vb{v} - \gamma_{\vb{v}} K \vb{v}$
        \STATE $\mathrm{rhs}_2 = -\lambda \vb{v} \vdot \idft{i \vb{k} \hat{\omega}} - \gamma_{\omega} K \omega$
        \STATE $\hat{\mathrm{rhs}} = i \vb{k} \times \vu{rhs}_1 + \hat{\mathrm{rhs}_2}$
        \STATE $\hat{\omega}_{\mathrm{naive}} = \exp\qty[\qty{a - \qty(1 - \vb{k}^2)^2}\Delta t]\qty(\hat{\omega} + \hat{\mathrm{rhs}} \Delta t )$
        \FORALL{$\vb{k}$}
        \IF{$\vb{k} = \vb{0}$}
        \STATE $\hat{\omega}_{\mathrm{new}} \qty(\vb{0}) = 0$
        \STATE $\vu{v}_{\mathrm{new}} \qty(\vb{0}) = \exp\qty[\qty(a - 1) \Delta t] \cdot \sum_{\vb{x}}\frac{1}{\sqrt{N^2}}(\vb{v} + \vb{rhs}_1 \Delta t)$
        \ELSIF{$\vb{k}^2 > \qty(\frac{1}{2} \cdot \frac{\pi}{\Delta x})^2$}
        \STATE $\hat{\omega}_{\mathrm{new}} \qty(\vb{k}) = 0$
        \STATE $\vu{v}_{\mathrm{new}} \qty(\vb{k}) = \vb{0}$
        \ELSE
        \STATE $\hat{\omega}_{\mathrm{new}}\qty(\vb{k})=\frac{1}{2}\qty{\hat{\omega}_{\mathrm{naive}}\qty(\vb{k})+\hat{\omega}_{\mathrm{naive}}\qty(\vb{-k})^*}$
        \STATE $\hat{\psi} \qty(\vb{k}) = \frac{\hat{\omega}_{\mathrm{new}} \qty(\vb{k})}{\vb{k}^2}$
        \STATE $\left. \vu{v}_{\mathrm{new}}\qty(\vb{k}) \right|_{x} = i \vb{k}_{y} \hat{\psi} \qty(\vb{k}),\ \left. \vu{v}_{\mathrm{new}} \qty(\vb{k}) \right|_{y} = - i \vb{k}_x \hat{\psi} \qty(\vb{k})$
        \ENDIF
        \ENDFOR
        \ENSURE{$\qty(\vb{v}_{\mathrm{new}},\ \hat{\omega}_{\mathrm{new}}) = \qty(\vb{v} \qty(t + \Delta t),\ \hat{\omega} \qty(t + \Delta t))$}
    \end{algorithmic}
    \caption{integration scheme} \label{alg:int}
\end{algorithm}

The key operations are the discrete Fourier transform (DFT) defined by $\hat{f} \qty(\vb{k}) = \frac{1}{\sqrt{N^2}} \sum_{\vb{x}} f \qty(\vb{x}) e^{-i \vb{k} \cdot \vb{x}}$ and its inverse (iDFT) similarly defined by $f \qty(\vb{x}) = \frac{1}{\sqrt{N^2}} \sum_{\vb{k}} \hat{f} \qty(\vb{k}) e^{i \vb{k} \cdot \vb{x}}$. Thanks to them, we can replace all the $\nabla$ by $-i \vb{k}$. Note that the periodic boundary condition is implicitly assumed, i.e., $f(x,\ y) = f(x \pm L,\ y) = f(x,\ y \pm L)$, whose artifacts are negligible as long as the walls are thick enough so that $\vb{v}(\vb{x})$ and $\omega(\vb{x})$ are sufficiently small at the periodic boundary.

As the first step, GPU calculates all the nonlinear terms using element-wise operations and DFT, which are then stored to $\hat{\mathrm{rhs}}$. A pitfall here is aliasing, i.e., artifacts caused by the expansion of the range of $\vb{k}$ due to multiplication in real space.
To deal with it, we need to eliminate high-wavenumber modes before and after this step, though the ``before'' operation can actually be optimized out.
For the TTSH equation containing the cubic term $b \vb{v}^2 \vb{v}$, the appropriate cutoff is $\frac{1}{2} \cdot \frac{\pi}{\Delta x}$ (the $\frac{1}{2}$ cutoff rule).

Next, GPU computes the time evolution of $\hat{\omega}$ by the Euler method and stores the result to $\hat{\omega}_{\mathrm{naive}}$. The replacement $\nabla \rightarrow -i \vb{k}$ plays a crucial role here, because it allows us to integrate the linear terms of the time evolution equation separately and exactly by using the exponential multiplier.
Without this technique, calculation will diverge unless we use ridiculously small $\Delta t \sim \Delta x^4$.

As the last step, GPU performs post-calculation operations.
For $\vb{k} = \vb{0}$, $\hat{\omega}$ is corrected to the exact value $0$ and $\vu{v}$ is directly calculated from the time evolution equation of $\vb{v}$, because this spatially uniform component cannot be recovered from $\omega$.
Specifically, we used the following equation, obtained by taking the spatial average of both sides of the TTSH equation [\eqref{eq:orig2}]:
\begin{equation}
    \pdv{\avgx{\vb{v}}}{t} = (a - 1) \avgx{\vb{v}} - b \avgx{\vb{v}^2 \vb{v}} - \gamma_{\vb{v}} \avgx{K\qty(\vb{x})\vb{v}}, \label{eq:avg}
\end{equation}
where the brackets denote spatial averaging. Note that $\grad{p}$, $\laplacian \vb{v}$, and $\vb{v} \vdot \grad{\vb{v}} = \div{\qty(\vb{v}\vb{v})}$ vanish because of the periodic boundary condition.
For $\vb{k}^2 > \qty(\frac{1}{2} \cdot \frac{\pi}{\Delta x})^2$, everything is set to $0$ to avoid aliasing.
For $0 < \vb{k}^2 \le \qty(\frac{1}{2} \cdot \frac{\pi}{\Delta x})^2$, $\hat{\omega}$ is corrected to satisfy $\hat{\omega} \qty(\vb{k}) = \hat{\omega} \qty(-\vb{k})^*$ and $\vb{v}$ is recovered from the corrected $\hat{\omega}$. The correction is essential for stable calculation; without that, accumulated numerical errors violate $\hat{\omega}_{\mathrm{naive}} \qty(\vb{k}) = \hat{\omega}_{\mathrm{naive}} \qty(-\vb{k})^*$, which is necessary for real-valued $\omega$.
For $\vb{v}$, GPU solves the equation for the stream function $\psi(\vb{x})$, $\laplacian{\psi} = -\omega$, and obtains $\vb{v}$ from $\psi$. This technique is valid for the two-dimensional incompressible flow studied here and $\div{\vb{v}} = 0$ is automatically satisfied. Without using this, one may need to solve Poisson's equation for the pressure, which is severely time-consuming, but here we can bypass it just by dividing $\hat{\psi}$ by $\vb{k}^2$.

\section{processing mask \texorpdfstring{$K(\vb{x})$}{K}} \label{sec:maskgen}
As described in the previous section, to avoid aliasing, it is necessary to apply a low-pass filter after real-space multiplications. This requirement also applies to the mask $K(\vb{x})$. Thus, we have to design an appropriate mask field $K(\vb{x})$ such that it has no high-wavelength modes, and at the same time it is nonnegative, $K \simeq 0$ inside the region of interest and $K \simeq 1$ otherwise. This can be automated as follows:
\begin{algorithm}[H]
    \begin{algorithmic}[1]
        \REQUIRE{$K_{\mathrm{naive}}$}
        \STATE $\hat{K}_{\mathrm{naive}} = \dft{K_{\mathrm{naive}}}$
        \FORALL{$\vb{k}$}
        \IF{$\vb{k}^2 > \qty(\frac{1}{4} \cdot \frac{\pi}{\Delta x})^2$}
        \STATE $\hat{K}_{\mathrm{naive}} \qty(\vb{k}) = 0$
        \ELSE
        \STATE modify $\hat{K}_{\mathrm{naive}} \qty(\vb{k})$ (optional)
        \ENDIF
        \ENDFOR
        \STATE $K = \idft{\hat{K}_{\mathrm{naive}}}^2$
        \ENSURE{processed $K(\vb{x})$}
    \end{algorithmic}
    \caption{mask processing scheme} \label{alg:mask}
\end{algorithm}

To begin with, \algref{alg:mask} computes DFT of $K_{\mathrm{naive}}$, followed by a low-pass filtering at a $\frac{1}{4}$ cutoff. This cutoff ensures that the output $K(\vb{x})$ does not violate the $\frac{1}{2}$ cutoff rule for antialiasing (we remark below why $\frac{1}{4}$).
Then, optionally, one may include additional low-pass operations (or any other processing) in the \textbf{else} block, as long as $\hat{K}_{\mathrm{naive}} \qty(\vb{k}) = \hat{K}_{\mathrm{naive}} \qty(-\vb{k})^*$ is satisfied. In the last step, $K$ is substituted by \textit{squared} iDFT of $\hat{K}_\mathrm{naive}$.
This ensures $K \ge 0$, without changing the essential shape of the mask.
It is because of this square that we chose the $\frac{1}{4}$ cutoff for $K_{\mathrm{naive}}$, to satisfy the $\frac{1}{2}$ rule in the finally obtained $K(\vb{x})$.

We used a binary mask for the input.
For the circular case, we set $K_{\mathrm{naive}} \qty(\vb{x}) = 1$ for $\norm{\vb{x}} \ge R$ and $0$ otherwise. For the dumbbell case, $K_{\mathrm{naive}}$ was obtained by logical AND of two circles.
We did not use any optional operations because any additional cutoff makes the difference between the designed geometry and the obtained $K(\vb{x})$ larger.

\section{why GPU?}  \label{sec:GPU}
The core observation is that any operation in the calculation scheme explained above reduces to an element-wise arithmetic operation either in the real or Fourier space. Element-wise operations can easily be parallelized, and using \texttt{cuFFT} library the DFT becomes massively faster. Note that any \texttt{if} statement should be purged from GPU codes for better performance, and in the present scheme it was possible to eliminate all the conditional branches from the performance-critical components.

\begin{figure}[tb]
    \centering
    \includegraphics[width = \hsize]{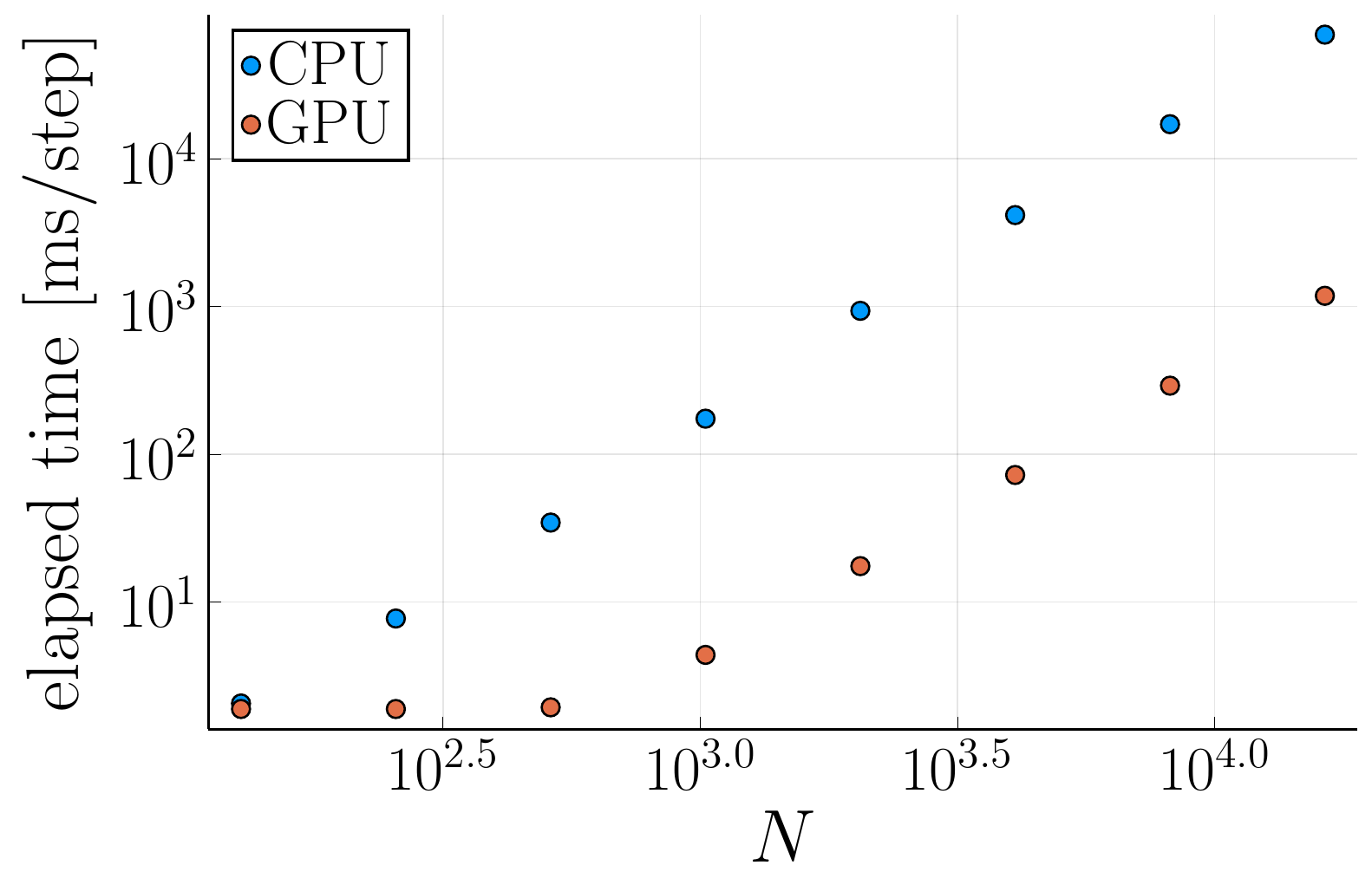}
    \caption{
        Time required for an iteration of the Euler method versus lattice size $N$, compared between CPU (Intel Xeon W-2295) and GPU (NVIDIA A6000). This shows that GPU is approximately $60$ times faster for large $N$.
    } \label{fig:time}
\end{figure}

As a result, although the computational cost of our large-scale simulations was far more massive compared to the previous studies, GPU was able to deal with them fast enough as shown in \figref{fig:time}.

\section{movie captions}

In \movrefs{S2-S8}, scales of the arrows for the velocity field and the colors for the vorticity field are kept unchanged. The color scale is identical to that used in \figref{fig:emoji_snap}.

\begin{description}
    \item[Movie S1] Vorticity field from a simulation with the emoji boundary. See \figref{fig:emoji} of the main text.
    \item[Movie S2] Vorticity fields for the circular confinement with radii $R=5.2,\ 5.4,\ \dots,\ 8.0$.
    \item[Movie S3] Velocity (arrows) and vorticity fields (color) for $R=5.2$ (single-vortex stationary state).
        This video includes the transient from a random initial condition.
    \item[Movie S4] Velocity (arrows) and vorticity fields (color) for $R=5.4$ (vortex-pair oscillatory state).
    \item[Movie S5] Velocity (arrows) and vorticity fields (color) for $R=5.6$ (chaotic state).
    \item[Movie S6] Velocity (arrows) and vorticity fields (color) for $R=5.8$ (quasiperiodic state).
    \item[Movie S7] Velocity (arrows) and vorticity fields (color) for $R=7.2$ (periodic state).
    \item[Movie S8] Velocity (arrows) and vorticity fields (color) for $R=7.6$ (chaotic state).
\end{description}

\section{code availability}
The codes used in this work and associated information are available upon request.

\begin{acknowledgments}
    We are grateful to Yusuke T. Maeda, Kazusa Beppu, Lailai Zhu for helpful discussions, Susumu Goto, Yuji Hattori, and Soshi Kawai for discussions from the perspective of fluid mechanics, Henning Reinken and Ryuna Nagayama for debugging, and Taichi Taniguchi for refactoring. The computation in this paper has been done partly by Kugui GPU supercomputer (ISSP, Univ.\ Tokyo) and nekoya/ai cluster (ipi, Univ.\ Tokyo). This work is supported in part by JSPS KAKENHI Grant Numbers JP19K23422, JP19H05800 and JP20K14426, and JST PRESTO Grant Number JPMJPR21O8, Japan.
\end{acknowledgments}

\bibliography{ref}

\end{document}